\documentclass[fleqn,usenatbib]{mnras}

\usepackage{newtxtext,newtxmath}

\usepackage[T1]{fontenc}

\DeclareRobustCommand{\VAN}[3]{#2}
\let\VANthebibliography\thebibliography
\def\thebibliography{\DeclareRobustCommand{\VAN}[3]{##3}\VANthebibliography}
\newcommand{\lta}{\lower 2pt \hbox{$\, \buildrel {\scriptstyle <}\over {\scriptstyle \sim}\,$}}
\newcommand{\gta}{\lower 2pt \hbox{$\, \buildrel {\scriptstyle >}\over {\scriptstyle \sim}\,$}}

\usepackage{graphicx}	
\usepackage{amsmath}	
\usepackage{subfigure}
\usepackage{caption}

\title[FRB Propagation]{Scattering of Strong Radio Waves by Particles in Strongly Magnetized Plasmas and Implications for Fast Radio Bursts}
\author[Y. Qu and P. Kumar]{
Yuanhong Qu$^{1,2}$\thanks{E-mail: yuanhong.qu@unlv.edu}
and
Pawan Kumar$^{3}$\thanks{E-mail: pk@astro.as.utexas.edu}\\
$^{1}$Nevada Center for Astrophysics, University of Nevada, Las Vegas, NV 89154\\
$^{2}$Department of Physics and Astronomy, University of Nevada Las Vegas, Las Vegas, NV 89154, USA\\
$^{3}$Department of Astronomy, University of Texas at Austin, Austin, TX 78712, USA
}

\date{}

\pubyear{2026}

\begin{document}
\label{firstpage}
\pagerange{\pageref{firstpage}--\pageref{lastpage}}
\maketitle

\begin{abstract}
Fast Radio Bursts (FRBs) are millisecond-duration radio transients that are widely believed to originate within magnetar magnetospheres. 
Large-amplitude radio waves associated with FRBs propagate through strongly magnetized plasmas, where nonlinear scattering can affect their propagation.
By solving the relativistic motion of a single particle interacting with electromagnetic waves of arbitrary polarization and propagation angle $\theta_B$, we compute the scattering cross section and the corresponding optical depth.
The scattering cross section of the O-mode can exceed that of the X-mode when $a\sin\theta_B < \omega_B/\omega$, and becomes comparable to that of the X-mode when $a\sin\theta_B > \omega_B/\omega$, where $\theta_B$ is the angle between the wave vector and the background field.
In the strongly magnetized and quasi-parallel limits, the cross sections asymptotically recover the linear regime scalings and are strongly suppressed by relativistic particle motion, leading to optical depths well below unity.
We also show that curvature radiation losses of O-mode waves are strongly suppressed for quasi-parallel propagation, allowing them to escape from the magnetosphere at moderate multiplicities.
We propose that Alfv\'en waves excited by magnetar crust quakes can reach amplitudes comparable to the background magnetic field, thus straightening field lines and reducing $\theta_B$.
This geometrical alignment enhances the ability of FRBs to freely propagate through the open field line region. 
These results suggest that large-amplitude waves propagating quasi-parallel to open magnetic field lines can avoid significant single-particle scattering losses, providing a possible condition for their escape.
\end{abstract}




\section{Introduction}

Fast Radio Bursts are intense, millisecond-duration radio signals with extremely high brightness temperature $\sim 10^{35} \ \rm K$ \citep{Lorimer2007,Thornton2013}.
Their extreme energetics and short durations point to compact, magnetized environments as the likely sources.
It is widely believed that at least some FRBs can be produced by magnetars since the detection of FRB 20200428 from a Galactic magnetar SGR 1935+2154 \citep{CHIME/FRB2020,Bochenek2020,CKLi21,Mereghetti20}.
The radiation mechanisms of FRBs are generally classified into two categories \citep{Zhang2020,ZhangRMP}: pulsar-like models, which invoke emission originating within or just outside the magnetar magnetosphere, and GRB-like models, which attribute the emission to relativistic magnetized shocks occurring far away from the central engine.
Recent observations of repeating FRBs strongly suggest that FRBs originate within the magnetosphere, based on signatures such as narrow spectra \citep{Pleunis2021,PKumar&Shannon2021,ZhangYK2023,ZhangLX2026}, nanosecond variability timescales \citep{Nimmo2022}, polarization angle (PA) swings and jumps \citep{Luo2020nature,NiuJR2024}, and significant circular polarization \citep{Jiang2024}. 
These diverse polarization states indicate that FRB emission consists of two orthogonal wave modes, both of which can be detected.
A non-repeater FRB 20221022A exhibited strong scintillation that constrained its source size, implying a magnetospheric origin \citep{Nimmo2025}. 
FRB 20221022A shows S-shaped PA \citep{Mckinven2024}, suggesting a similar magnetospheric emission between FRBs and pulsar radio emission. 
Recent theoretical energetic constraints further support magnetospheric models \citep{Beniamini&Kumar2026}.

Most observations imply a magnetospheric origin of FRBs. 
However, the physical conditions that enable strong FRB radio waves to escape the magnetosphere remain an open problem.
In general, these issues are related to the amplitude and frequency of FRBs:
\begin{itemize}
\item In the high frequency regime where the typical frequency of FRBs is much higher than the plasma frequency, i.e. $\omega\gg\omega_p$.
X- and O-mode waves can exist and propagate.
The X-mode is polarized perpendicularly to the $\vec k-\vec B$ plane.
The electric field of the O-mode wave is polarized in the $\vec k-\vec B$ plane. 
Theoretical challenges arise from the strong electromagnetic field of an FRB pulse, characterized by a large amplitude parameter $a=eE_w/(m_ec\omega)\gg1$, where $E_w$ denotes the wave amplitude.
The strength of FRB waves can exceed the background magnetic field near the light cylinder of the magnetar and the waves enter the non-linear scattering regime.
The scattering cross section of electrons in the background magnetic field can far exceed the Thomson value, suggesting that the waves might be scattered before leaving the magnetosphere \citep{Beloborodov2021,Beloborodov2022}.
However, this non-linear scattering problem can be mitigated if the radiation propagates along the open field line region, where the plasma streams outward with a high Lorentz factor $\gamma\gg1$, and the angle $\theta_B$ between the FRB wave vector and the background magnetic field is small \citep{QKZ,Lyutikov2024,Huang2024}.
These effects reduce the scattering cross section and allow luminous FRBs to escape in the open field line region.
\item  In the low frequency regime where $\omega\ll\omega_p$.
The X-mode becomes fast magnetosonic (fast) mode and O-mode becomes Alfv\'en waves.
The X-mode and fast mode waves have the same polarization with the electric field perpendicular to the ($\vec k - \vec B$) plane and the same dispersion relation ($\omega=kc$) in the highly magnetized pair plasma \citep{Stix1992}. 
The difference is that the X-mode is defined above the plasma frequency whereas the fast mode is defined below the plasma frequency.
The Alfv\'en waves can only propagate along the background magnetic field because the parallel electric field is rapidly screened by plasma oscillations.
In this regime, fast mode waves can be converted into Alfv\'en waves and therefore cannot escape from the magnetosphere \citep{Golbraikh&Lyubarsky2023}.
Specifically, large-amplitude fast mode can become nonlinear and steepen into shocks in the closed field line region \citep{ChenYR2022,Beloborodov2023,Vanthieghem&Levinson2025,Bernardi2025}.
These physical processes can further constrain the emission regions of FRBs.
\end{itemize}

The propagation of FRBs is related to the background magnetic field geometry, which is unlikely to be stable in FRB-emitting magnetar magnetospheres. 
Observations of PA swings suggest that the magnetosphere of the FRB central engine is continuously distorted by the emission process and that the magnetic configuration is dynamically evolving \citep{Liu2025}. 
The background field is likely perturbed by magnetohydrodynamic (MHD) waves, including Alfv\'en waves and fast mode waves, which are excited by crustal quakes near the magnetar surface \citep{Qu&Bransgrove}. 
As Alfv\'en waves propagate outward, the background magnetic field can be progressively opened and become more aligned with the FRB propagation direction \citep{QKZ}. 
The magnetar quakes can be coupled with detailed force-free electrodynamic simulations of the magnetosphere. 
In this regime the magnetic field evolves into a strongly perturbed split monopole configuration after entering the nonlinear radius $R_{\rm E,aw}$ where $B_w=B$ \citep{Burnaz2025}.

\begin{figure*}
\centering
\includegraphics[width=17cm,height=6cm]{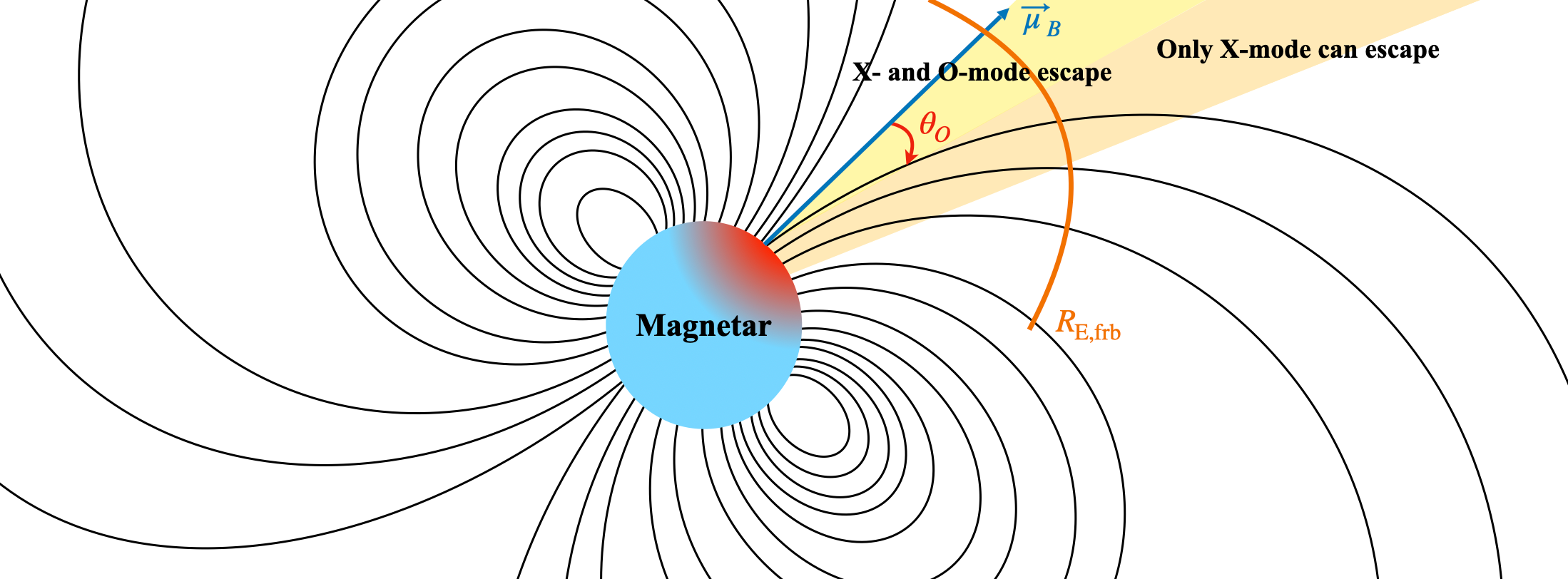}\\[4pt]
\includegraphics[width=17cm,height=6cm]{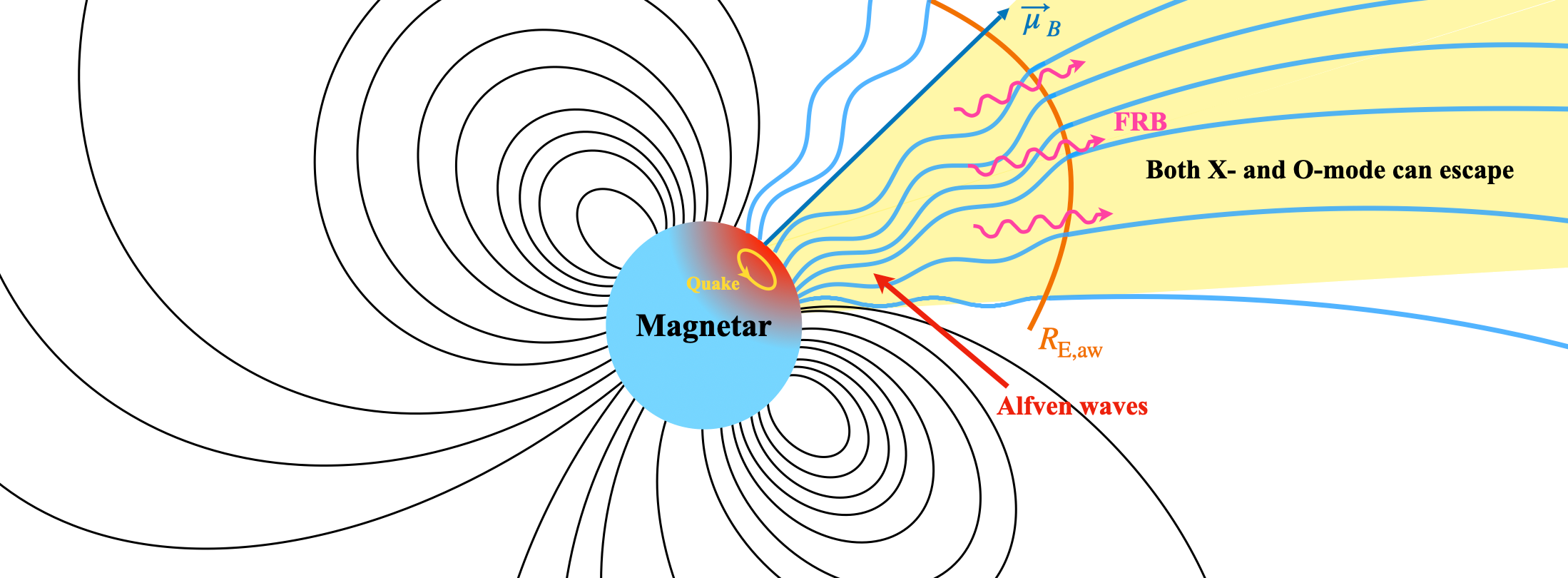}
\caption{Top panel: In a static dipolar magnetic field, both X- and O-mode waves can escape from the magnetosphere in regions close to the magnetic moment, where the wave vector is nearly quasi-parallel to the background magnetic field (yellow region). 
In regions farther from the magnetic moment (orange region), only X-mode waves can escape, since O-mode waves undergo stronger nonlinear scattering when propagating obliquely at radii $r > R_{\rm E,frb}$.
Bottom panel: The background magnetic field is dynamically evolved. 
Alfv\'en waves (blue wiggler) can be excited by magnetar crustal quakes near the stellar surface. 
As they propagate outward, the Alfv\'en waves become nonlinear at $R_{\rm E,aw}$, causing the background dipolar magnetic field to open up and approach a monopolar configuration. 
In this quasi-parallel geometry, both X- and O-mode waves are able to escape.}
\label{fig:cartoon}
\end{figure*}

The main components of the scenario we explore in this work are described in Figure~\ref{fig:cartoon}. 
We first investigate O-mode wave propagation in the linear regime inside the magnetar magnetosphere, then show analytically the escape condition.
We explore the non-linear scattering cross section of X-mode, O-mode, mixed mode and circular polarized waves numerically.
We incorporate the effect of non-linear Alfv\'en waves excited during magnetar bursts. 
As these Alfv\'en waves evolve into the non-linear regime, they can distort and straighten magnetic field lines, effectively reducing the local $\theta_B$.
This geometrical alignment dynamically enhances the transparency of the open field line region, allowing FRB waves to propagate with reduced attenuation (see the lower panel of Figure~\ref{fig:cartoon}).
A rotating polar-cap scenario has been proposed to connect the repetition rate and polarization properties of FRBs to the alignment between the magnetic and rotation axes, further motivating models in which radio emission is produced in the open magnetic field line region \citep{Beniamini&Kumar2025}.

The paper is organized as follows. 
In Section~\ref{sec:Particle motion}, we perform a detailed numerical calculation of the scattering cross section for a test particle, initially at rest, interacting with FRB waves of mixed linear modes and circular polarization.
In Section~\ref{sec:Scattering}, we investigate the O-mode damping process in the linear regime and present numerical results for scattering cross sections of X-mode, O-mode, and mixed mode in Section~\ref{sec:Numerical results}.
In Section~\ref{sec:open}, we discuss that the background magnetic fields open up when Alfv\'en waves become non-linear and its observational implications.
The main conclusions and  discussions are presented in Section~\ref{sec:Conclusion}.

\section{Particle motion in a large-amplitude wave and a strong background magnetic field}\label{sec:Particle motion}

In this section, we present a general theoretical framework to describe the motion of a single charged particle in the presence of a background magnetic field and a large-amplitude electromagnetic wave.
The dimensionless non-linear parameter that characterizes the strength of an electromagnetic wave in vacuum is defined as
\begin{equation}
a=\frac{eE_w}{m_ec\omega},
\end{equation}
where $\omega$ and $E_w$ denote the angular frequency and electric field strength of FRB waves, respectively. 
We note that $a$ is Lorentz invariant.
Particles can be accelerated to relativistic speeds approaching the speed of light in vacuum over a characteristic timescale of order $\sim\omega^{-1}$ when $a>1$.
In the presence of a background magnetic field, the field strength $B$, the wave propagation angle $\theta_B$, and the wave polarization all enter the problem and result in more complex particle dynamics.

To describe the motion of a particle in a large-amplitude FRB wave and a background magnetic field, it is convenient to work with the Lagrangian for a test particle, which is given by \citep{Jackson1998}
\begin{equation}
L=-\frac{m_ec^2}{\gamma}-q\phi+\frac{q}{c}\vec A\cdot\vec v,
\end{equation}
and the canonical momentum
\begin{equation}
\vec P=\gamma m_e\vec v+\frac{q}{c}\vec A,
\end{equation}
where $\gamma$ is the Lorentz factor of the electron, $\phi$ is the electric potential associated with the FRB wave, and $\vec A$ is the vector potential associated with both the FRB wave and the background magnetic field $B$.
Following \cite{QKZ}, we consider the general case of a background magnetic field oriented at an arbitrary angle $\theta_B$ with respect to the FRB wave vector.
In the spherical coordinate system with the $z$-axis defined along the wave vector direction, the unit vector of the background magnetic field is $\hat{e}_{\rm bg}=(\sin\theta_B\cos\phi_B,\ \sin\theta_B\sin\phi_B,\ \cos\theta_B)$, where $\phi_B$ is the azimuthal angle of $\vec B$.
We consider two different vector potentials for the background magnetic field in order to evaluate the canonical momenta
\begin{equation}\label{eq:A_0_1}
\vec A_1=B_yz\hat{x}-B_xz\hat{y}+B_zx\hat{y},
\end{equation}
and
\begin{equation}\label{eq:A_0_2}
\vec A_2=B_yz\hat{x}-B_zy\hat{x}-B_xz\hat{y},
\end{equation}
where $B_x=B\sin\theta_B\cos\phi_B$, $B_y=B\sin\theta_B\sin\phi_B$ and $B_z=B\cos\theta_B$ denote the three components of the background magnetic field.
In the following, we discuss the linear and circular polarization states of the wave in Section~\ref{subsec:linear polarized} and \ref{subsec:circular polarized}. In Section~\ref{subsec:cross section} we calculate the cross section.

\subsection{Particle Dynamics in Linearly Polarized Waves}\label{subsec:linear polarized}

For a fully linearly polarized wave, we consider the EM wave propagating along the $z$-axis and the corresponding electric and magnetic fields are
\begin{equation}
\vec E_w=E_w\sin\psi\hat{x}, \ \vec B_w=E_w\sin\psi\hat{y},
\end{equation}
where $\psi=kz-\omega t$ is the phase.
The corresponding vector potential of the EM wave can be written as
\begin{equation}
\vec A_w=-\frac{c}{\omega}E_w\cos\psi\hat{x}, \ \psi=kz-\omega t.
\end{equation}
Using the conserved canonical momenta in the three coordinate directions, one can derive the equations of motion \citep{QKZ}
\begin{equation}
\gamma v_x=ac(\cos\psi-\cos\psi_0)-\omega_B(\sin\theta_B\sin\phi_B z-\cos\theta_By).
\end{equation}
\begin{equation}
\gamma v_y=\omega_B(\sin\theta_B\cos\phi_B z-\cos\theta_B x),
\end{equation}
\begin{equation}
\gamma v_z=(\gamma-1)c+\omega_B\sin\theta_B(x\sin\phi_B-y\cos\phi_B),
\end{equation}
where $\omega_B=eB/(m_ec)$ is the cyclotron frequency, we apply the initial conditions as $x=y=z=0$, $\psi=\psi_0$ and $\vec v=(v_x,v_y,v_z)=0$ at $t=0$, i.e. the particle is at rest before it encounters the wave.
The kinetic energy equation of the particle can be written as
\begin{equation}
\frac{d(\gamma m_ec^2)}{dt}=qv_xE_w\sin\psi.
\end{equation}
The Lorentz factor can be calculated through
\begin{equation}
\begin{aligned}
&2\gamma[1-\omega_B\sin\theta_B(x\sin\phi_B - y\cos\phi_B)/c]=2+\frac{\omega_B^2}{c^2}\times\\
&(\sin\theta_B\cos\phi_Bz-\cos\theta_Bx)^2 + \frac{\omega_B^2 \sin^2\theta_B}{c^2}(x \sin\phi_B - y \cos\phi_B)^2 \\
& - 2\omega_B\sin\theta_B(x\sin\phi_B-y\cos\phi_B)/c\\
&+[a(\cos\psi-\cos\psi_0)-\omega_B(\sin\theta_B\sin\phi_Bz-\cos\theta_By)/c]^2.
\end{aligned}
\end{equation}
One can see that $\gamma=1+a^2(\cos\psi-\cos\psi_0)^2/2$ when $B=0$, and the Lorentz factor along the wave vector is $\gamma_z\sim a$.
The above analysis is valid for relativistic particles with $v\rightarrow c$ at $t=0$ and is most conveniently carried out in the comoving frame.
The initial polarization state of a fully linearly polarized wave is determined by $\phi_B$. 
The X-mode corresponds to $\phi_B=\pi/2$, the O-mode corresponds to $\phi_B=0$, and a fully linearly polarized mixed mode corresponds to $0<\phi_B<\pi/2$.

\subsection{Particle Dynamics in Circularly Polarized Waves}\label{subsec:circular polarized}

For a fully circularly polarized wave, the electric field can be decomposed into two orthogonal linearly polarized components as
\begin{equation}
\vec E_{w1}=E_w\sin\psi\hat{x},
\end{equation}
and
\begin{equation}
\vec E_{w2}=E_w\sin\left(\psi+\frac{\pi}{2}\right)\hat{y}=E_w\cos\psi\hat{y}.
\end{equation}
The total electric field is related to the vector potential through $\vec E_w=-\partial \vec A_w/(c\partial t)$, from which the vector potential of the wave can be written as
\begin{equation}
\vec A_w=-\frac{c}{\omega}E_w\cos\psi\hat{x}+\frac{c}{\omega}E_w\sin\psi\hat{y}.
\end{equation}
The motion equations can be derived as
\begin{equation}
\gamma v_x=ac(\cos\psi-\cos\psi_0)-\omega_B(\sin\theta_B\sin\phi_Bz-\cos\theta_By),
\end{equation}
\begin{equation}
\gamma v_y=\omega_B(\sin\theta_B\cos\phi_Bz-\cos\theta_Bx)+ac(\sin\psi_0-\sin\psi),
\end{equation}
\begin{equation}
\gamma v_z=c(\gamma-1)+\omega_B\sin\theta_B(x\sin\phi_B-y\cos\phi_B),
\end{equation}
and the kinetic energy equation of the particle can be written as
\begin{equation}
\frac{d(\gamma m_ec^2)}{dt}=q(v_xE_w\sin\psi+v_yE_w\cos\psi).
\end{equation}
The Lorentz factor can be calculated through
\begin{equation}
\begin{aligned}
&2\gamma[1-{\omega_B\sin\theta_B}(x\sin\phi_B-y\cos\phi_B)/c]=2
\\
&+{\omega_B^2\sin^2\theta_B}(x\sin\phi_B-y\cos\phi_B)^2/{c^2}\\
&-2{\omega_B\sin\theta_B}(x\sin\phi_B-y\cos\phi_B)/c\\
&+\left[a(\cos\psi-\cos\psi_0)-{\omega_B}(\sin\theta_B\sin\phi_Bz-\cos\theta_By)/c\right]^2
\\
&+\left[a(\sin\psi_0-\sin\psi)+{\omega_B}(\sin\theta_B\cos\phi_Bz-\cos\theta_Bx)/c\right]^2 .
\end{aligned}
\end{equation}
The Lorentz factor is $\gamma=1+a^2(1-\cos\psi)$ when $B=0$ and $\psi_0=0$, the instantaneous Lorentz factor in this case is always larger than that in the linearly polarized case.
We note that in the circularly polarized case the expressions for $\gamma v_x$ and $\gamma v_z$ remain identical to those obtained for linear polarization, while only the transverse component $\gamma v_y$ is modified.
Circular polarization introduces an additional $y$-component of the vector potential, altering the conserved quantity $P_y$ and therefore modifying only the transverse motion in the plane perpendicular to the propagation direction.

\subsection{Scattering cross section of a relativistic particle}\label{subsec:cross section}

The trajectory of the particle can be calculated using the first order differential motion equations, and the emitted power of a relativistic charged particle can be calculated as \citep{Jackson1998} 
\begin{equation}
\begin{aligned}
P&=\frac{2q^2}{3 c^3} \sum_{\alpha=0}^3 \frac{d(u^\alpha)}{ d\tau} \frac{d(u_\alpha)}{d\tau}\\
&=\frac{2q^2\gamma^2}{3c^3}\left[-\left(\frac{du^0}{dt}\right)^2+\sum_{i=x,y,z}\left(\frac{du^i}{dt}\right)^2\right],
\end{aligned}
\end{equation}
where $u^\alpha=\gamma(c,v_x,v_y,v_z)$ is the four-velocity and $d\tau=dt/\gamma$ is the differential proper time.
The relation between the cross sections in the comoving and lab frames follows
\begin{equation}
\sigma'=\frac{P'}{S'}=\frac{\sigma}{1-\beta\cos\theta_B},
\end{equation}
where $\sigma$ is the cross section in the lab frame, $P'$ and $S'$ denote the emitted power of the particle and Poynting flux of incident waves in the comoving frame. 
We first solve the particle dynamics in the comoving frame, then evaluate the radiated power and the resulting scattering cross section, and finally transform the cross section to the laboratory frame.
When the wave propagates in the closed field line region of the magnetar with $\theta_B\lesssim 90^\circ$, the cross section in the lab frame is comparable to that in the comoving frame, i.e. $\sigma\lesssim \sigma'$.
In contrast, when the wave is propagating quasi-parallel to the background magnetic field, i.e. when the local field lines are open where $\theta_B\ll1$, the cross section is reduced according to
\begin{equation}
\sigma\simeq (1-\beta)\sigma'\simeq \frac{\sigma'}{2\gamma^2},
\end{equation}
where the last step applies for relativistic particles with the initial Lorentz factor $\gamma\gg1$ before interacting with the incoming wave.
One can see that the cross section will be highly suppressed when waves propagate through the open field line region.
In this case the wave catches the relativistic particle and the cross section is reduced.

\begin{figure}
\hspace*{-2mm}
\includegraphics[width=86mm]{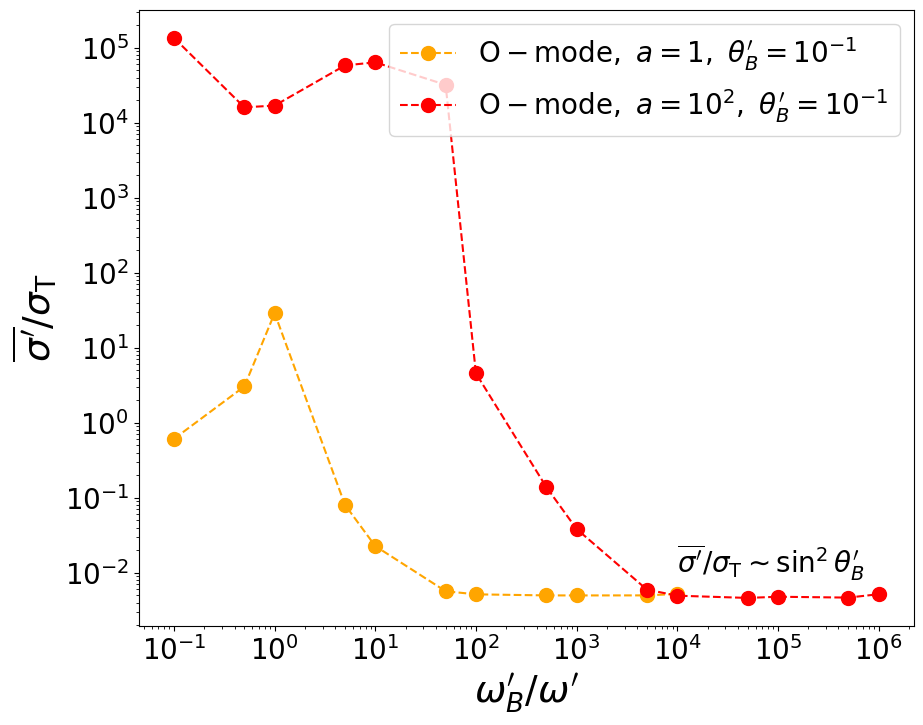}
    \caption{The normalized cross section $\overline{\sigma'}/\sigma_{\rm T}$ as a function of $\omega'_B / \omega'$ for O-mode waves with $a=1$ (orange) and $a=10^2$ (red), evaluated at a fixed propagation angle $\theta_B'=0.1$.}
\label{fig:sigma_Omode}
\end{figure}

In the linear regime where $E_w\ll B$, the cross sections of X- and O-mode waves in the comoving frame are given by \citep{Herold1979}
\begin{equation}
\sigma'({\rm X})=\frac{\sigma_{\rm T}}{2}\left[\frac{\omega'^2}{(\omega'+\omega_B')^2}+\frac{\omega'^2}{(\omega'-\omega_B')^2}\right],
\label{eq:Xmode}
\end{equation}
and
\begin{equation}
\begin{aligned}
\sigma'({\rm O})&=\sigma_{\rm T}\sin^2\theta_B'+\sigma'({\rm X})\cos^2\theta_B',
\end{aligned}
\end{equation}
where $\sigma_{\rm T}$ is the Thomson cross section.
The cross section of X-mode waves still follows $\sigma'({\rm X})\simeq \sigma_{\rm T}(\omega'/\omega_B')^2$ when $a\sin\theta_B<\omega_B/\omega$ \citep{QKZ}.
Here we calculate the cross section of O-mode waves with $a=1$ and $10^2$ as a function of the ratio of the cyclotron and FRB angular frequencies, i.e. $\omega_B'/\omega'$ in Figure~\ref{fig:sigma_Omode}.
For a given set of parameters, we run the simulations long enough to derive a reliable average value $\bar\sigma'$ for a time duration $t\simeq 1 \ \rm ms$, which is the typical time duration of FRBs.
One can see that $\sigma'({\rm O})\sim \sigma_{\rm T}\sin^2\theta_B'$ when $\omega_B'/\omega'\gg1$.
Thus we conclude that the two linear expressions can still be valid in the highly magnetized plasma even for $a\gg1$.

The numerical results of cross section are presented in Section~\ref{sec:Numerical results} for extremely large-amplitude waves with $a\gg1$ and arbitrary polarization states cases.

\section{O-mode waves propagation in the linear regime}\label{sec:Scattering}

For a dipolar magnetic field with $B_\star=10^{15} \ \rm G$ at the surface of the magnetar, the transition radius where $E_w=B$ can be estimated as
\begin{equation}
R_{\rm E,frb}=\left( \frac{B_\star R_{\star}^3 c^{1/2}}{L_{\rm frb}^{1/2}} \right)^{1/2} = (7.7\times10^8\,{\rm cm})\; B_{\star,15}^{1/2} R_{\star,6}^{3/2} L_{\rm frb,42}^{-1/4}.
\end{equation}
where $L_{\rm frb}$ denotes the typical luminosity of FRBs.
The corresponding nonlinear parameter of the FRB waves evaluated at $R_{\rm E,frb}$ is
\begin{equation}\label{eq:typical_a_parameter}
a\simeq 1.6\times 10^4 \ L_{\rm frb,42}^{1/2}\nu_9^{-1}r_9^{-1}\gg1,
\end{equation}
where $\nu=1 \ \rm GHz$ is the typical frequency of FRBs.

In this section, we focus on the propagation of large-amplitude O-mode waves in the linear regime where $E_w<B$. 
We first investigate particle acceleration by the parallel electric field of O-mode waves, the resulting curvature radiation, and the possible onset of pair production in Section~\ref{subsec:O-mode_curvature}.
We then discuss the scattering optical depth of O-mode waves in Section~\ref{subsec:O-mode_linear_scattering}.

\subsection{Curvature Radiation and Pair Production}\label{subsec:O-mode_curvature}

\begin{figure*}
\begin{center}
\setlength{\tabcolsep}{-12pt}
\begin{tabular}{ll}
\resizebox{95mm}{!}{\includegraphics[]{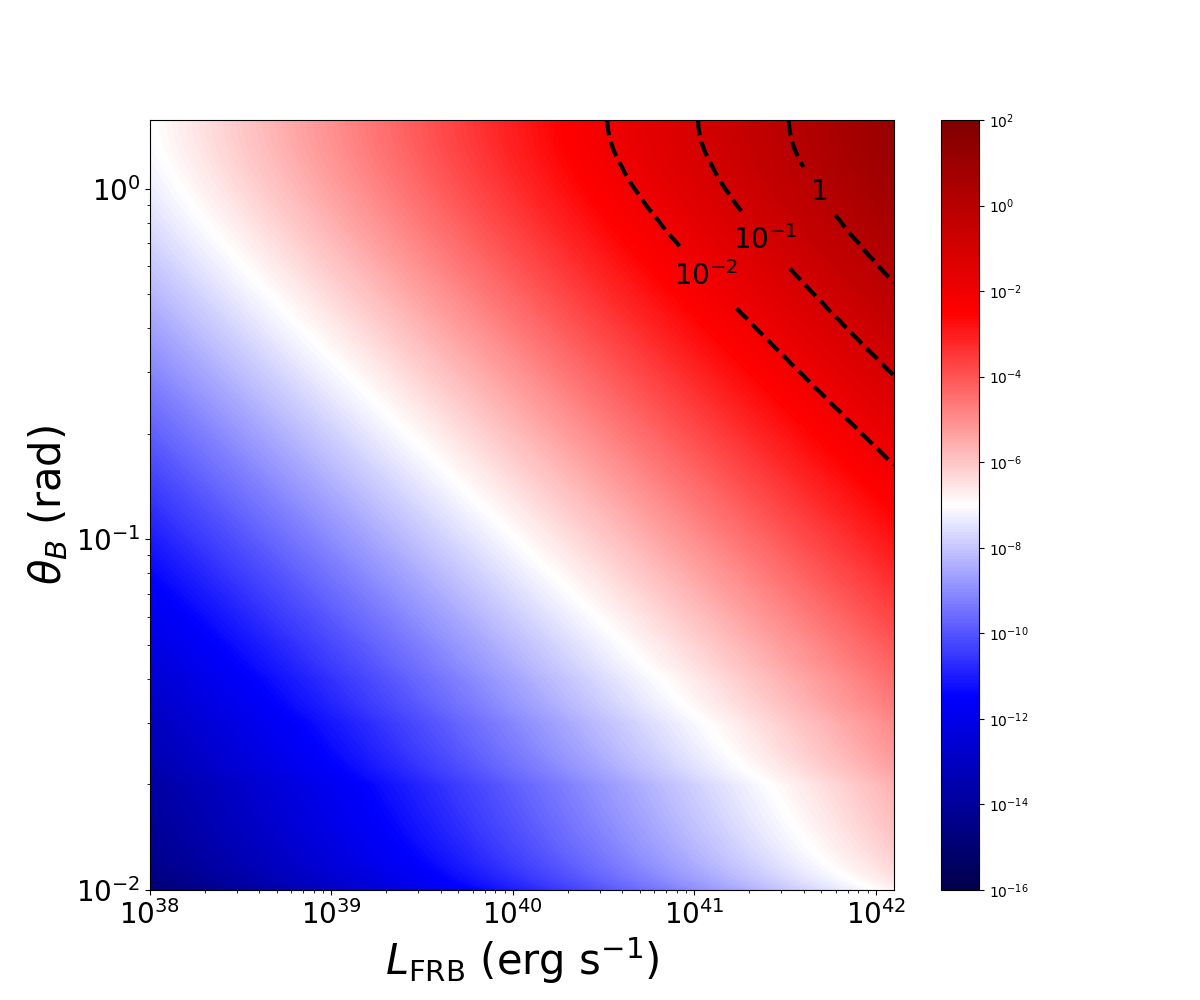}}&
\resizebox{95mm}{!}{\includegraphics[]{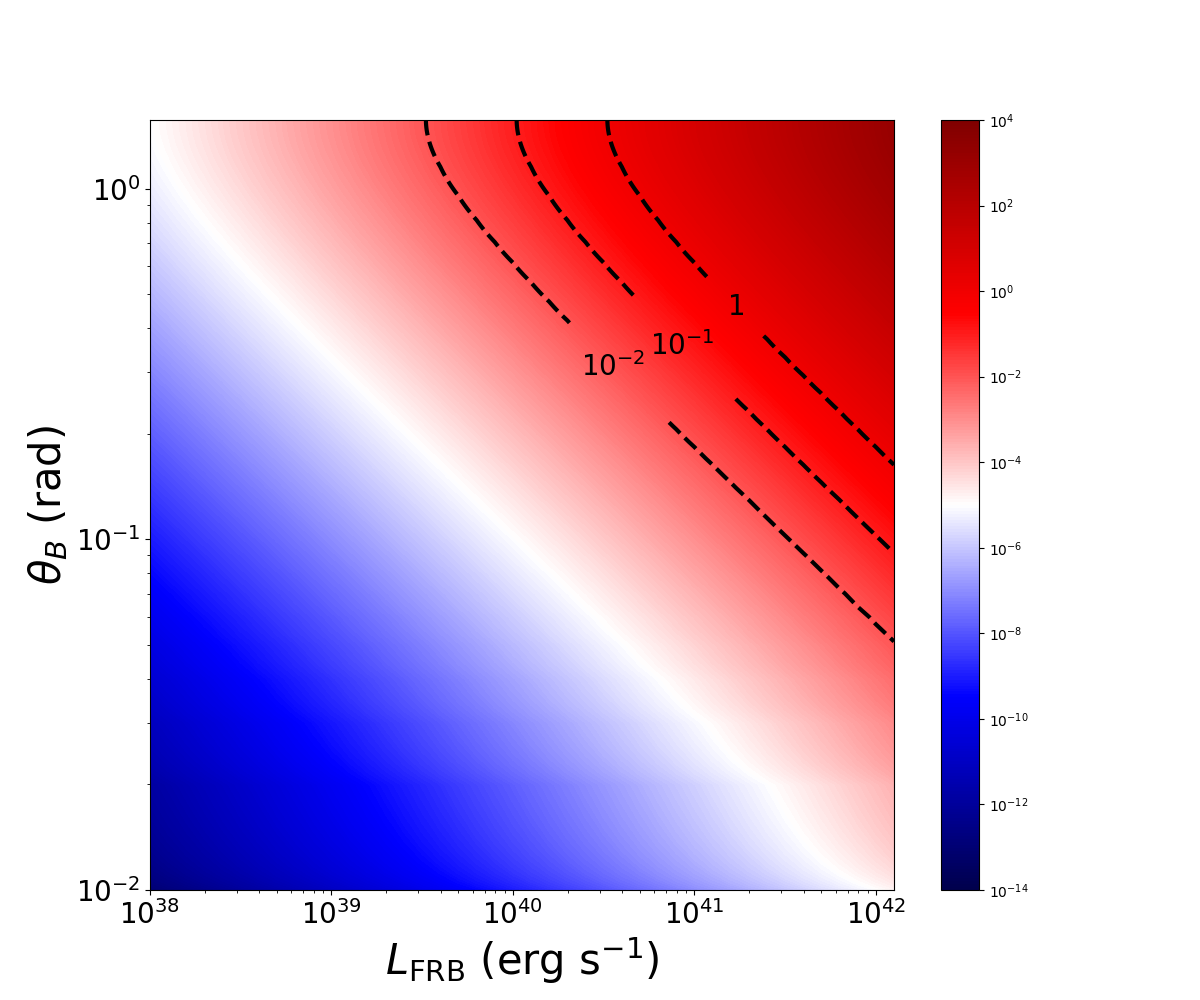}}\\
\resizebox{95mm}{!}{\includegraphics[]{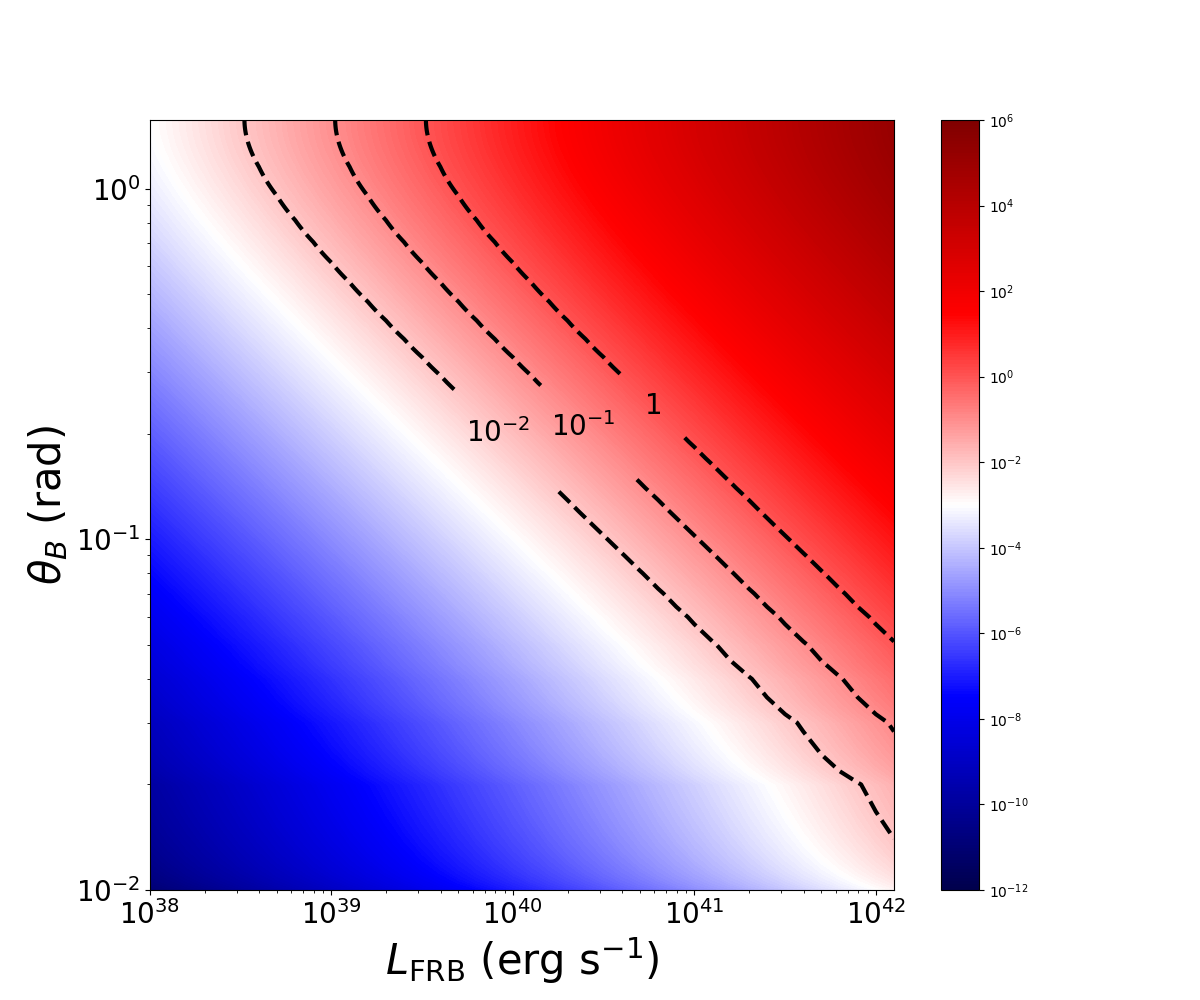}}&
\resizebox{95mm}{!}{\includegraphics[]{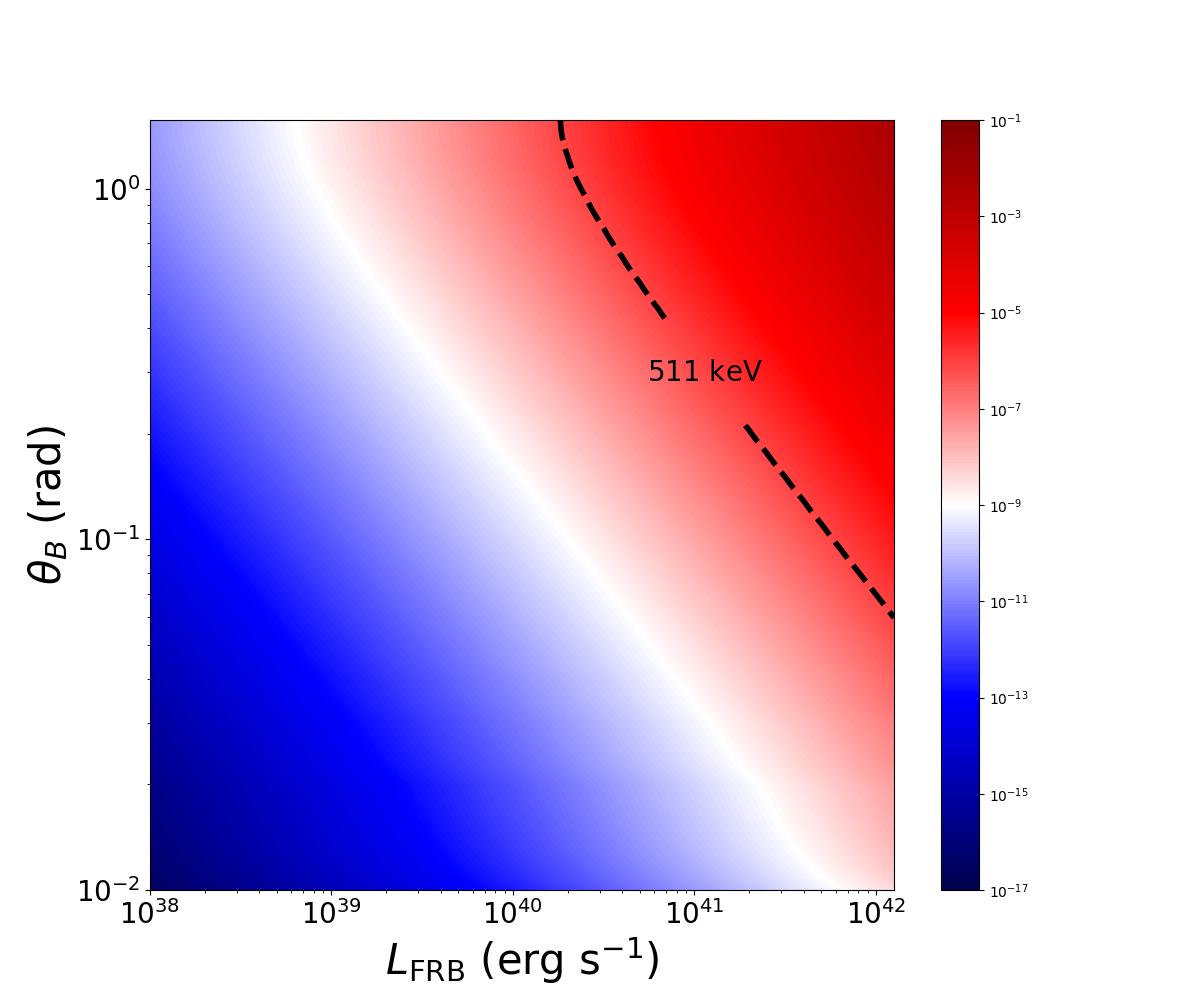}}
\end{tabular}
\caption{Total emitted curvature-radiation energy (color scale; upper left, upper right, and lower left panels) and characteristic energy of curvature photons (lower right panel) as functions of the FRB luminosity $L_{\rm FRB}$ for O-mode waves and the propagation angle $\theta_B$.
The upper left, upper right, and lower left panels correspond to $\xi=10^2$, $10^4$, and $10^6$, respectively.
The three black dashed contours indicate the fractional energy loss due to curvature radiation, defined as $f_{\rm cur}=E_{\rm cur}/E_{\rm frb}$, with levels $10^{-2}$, $10^{-1}$, and $1$.
The black dashed line in the lower right panel marks the location where the characteristic photon energy reaches $m_e c^2 = 511 \ \rm keV$.
Following parameters are adopted: FRB time duration $\Delta t_{\rm frb}=1 \ \rm ms$, magnetar spin $P=1 \ \rm s$, magnetar surface magnetic field $B_\star=10^{15} \ \rm G$ and radius $R_\star=10^6 \ \rm cm$.}
\label{fig:contour_Omode_curvature}
\end{center}
\end{figure*}

Since the electric field of O-mode waves lies in the $(\vec k -\vec B)$ plane, an unscreened parallel electric field $E_\parallel=E_w\sin\theta_B$ arises when the wave propagates at a finite angle to the magnetic field, i.e. for $\theta_B>0$. 
As a result, O-mode waves may suffer energy losses because charged particles can be accelerated by the parallel electric field and subsequently emit curvature radiation.

We consider a particle initially at rest. 
The Lorentz factor associated with particle motion along the background magnetic field can reach $\gamma_\parallel\approx a\sin\theta_B$.
The particle could attain relativistic speed when $a\gg (\sin\theta_B)^{-1}$ and can radiate efficiently due to acceleration induced by the O-mode waves.
The particle predominantly moves along the strong background magnetic field $B$.
However, the presence of the magnetic field $B_w$ of O-mode waves leads to a slight tilt of the local magnetic field direction.
Consider $E_w\simeq B_w\ll B$ at $r\ll R_{\rm E,frb}$ and the corresponding tilt angle defined as the instantaneous deflection angle of the particle momentum direction can be estimated as $\theta=\arctan\left({B_w}/{B}\right)\simeq {E_w}/{B}$.
The characteristic curvature radius of the particle trajectory is then given by $r_c\simeq {c}/{(\theta\omega)}\simeq{cB}/{(\omega E_w)}$.
The curvature radiation power emitted by a single particle can be estimated as
\begin{equation}
P_{\rm cur}=\frac{2e^2c\gamma_\parallel^4}{3r_c^2}\simeq \frac{2e^2}{3c}a^4\omega^2 \frac{E_w^2}{B^2}\sin^4\theta_B.
\end{equation}
The total energy radiated via curvature emission over the FRB time duration $\Delta t_{\rm frb}$ can be written as
\begin{equation}
E_{\rm cur}\simeq \frac{2e^2}{3c}a^4\omega^2 \frac{E_w^2}{B^2}\sin^4\theta_B\Delta t_{\rm frb}N_{\rm tot},
\end{equation}
where $N_{\rm tot}\sim\xi n_{\rm GJ}V$ and $V\sim\pi r^3$ denote the total number of particles and volume, respectively.
$\xi$ is the multiplicity factor\footnote{The plasma multiplicity in FRB-emitting magnetars is highly uncertain. Observational constraints from diverse polarization angle swings and significant circular polarization place limits on the background plasma density \citep{QZK2026}.
If the multiplicity is too large, only a single wave mode (the fast mode) can escape from the magnetosphere, which is inconsistent with current FRB polarization observations.} and $n_{\rm GJ}$ is the Goldreich-Julian number density \citep{Goldreich&Julian1969}.
The total energy of the FRB at radius $r$ can be estimated as $E_{\rm frb}\simeq (E_w^2/4\pi)4\pi r^3/3= E_w^2r^3/3$.
We define the fractional energy loss due to curvature radiation as $f_{\rm cur}=E_{\rm cur}/E_{\rm frb}$.
We present the curvature radiation energy loss fraction as a function of the FRB luminosity $L_{\rm FRB}$ and the propagation angle $\theta_B$ in the upper left, upper right and lower left panels of Figure~\ref{fig:contour_Omode_curvature} for $\xi=10^2, 10^4$ and $10^6$, respectively.
The three black dashed contours correspond to $f_{\rm cur}=10^{-2}$, $10^{-1}$ and $1$.
For $\xi=10^2$, the energy loss due to curvature radiation remains below the percent level as long as $\theta_B\lesssim 0.1$.
We also extend the calculation to higher multiplicities and find that the curvature radiation energy loss becomes significantly enhanced in a substantial region of parameter space. 
This implies that a large fraction of the FRB energy would be dissipated during propagation through the magnetosphere.
In addition, a higher pair multiplicity also increases the plasma density, leading to a larger optical depth for wave propagation. 
As a result, for sufficiently large $\xi$, FRB pulses may not be able to escape from magnetospheres.
This suggests that magnetars with extremely high pair multiplicities are unlikely to serve as viable sources of magnetospheric FRBs.

The characteristic angular frequency of the curvature radiation is given by $\omega_c={3c\gamma_\parallel^3}/{2 r_c}$.
The characteristic energy of the emitted curvature photons as a function of the FRB luminosity $L_{\rm FRB}$ and the propagation angle $\theta_B$ is shown in the lower right panel of Figure~\ref{fig:contour_Omode_curvature}. 
The black dashed line indicates where the photon energy reaches $m_ec^2=511 \ \rm keV$. 
Below this line, pair production is not possible. 
This region corresponds to small propagation angles $\theta_B \lesssim 0.1$.
Curvature emission and the associated pair production are therefore strongly suppressed when O-mode waves propagate quasi-parallel to the local magnetic field in the linear regime. This behavior is expected because curvature photons are highly beamed along the direction of motion of the relativistic particles. 
In addition, the Alfv\'en waves could open up the magnetosphere and lead to a more aligned geometry between $\vec k$ and $\vec B$ prior to FRB emission (see Section~\ref{sec:open} and Equation~(\ref{eq:R_E_aw})).
The particles are likely initially relativistically moving with a bulk Lorentz factor $\Gamma$, the pair production threshold condition is modified to $(\hbar\omega_c)^2\gtrsim \Gamma^2(m_ec^2)^2$, which effectively raises the required photon energy by a factor of $\Gamma$.
This further suppresses pair production induced by curvature radiation from O-mode waves.

\subsection{Scattering of O-mode waves}\label{subsec:O-mode_linear_scattering}

In a realistic open field line region of the magnetar magnetosphere, charged particles are expected to be already relativistically moving before interacting with FRB waves.
In this regime, particle inertia strongly suppresses rapid changes in the particle trajectory, and the wave -- particle interaction is therefore described in terms of scattering rather than curvature radiation.
In the following, we investigate whether scattering can become an efficient damping mechanism for O-mode FRB waves.

The background magnetic field can be regarded as locally straight compared to the typical FRB wavelength.
Charged particles therefore propagate nearly along the background magnetic field, and the cyclotron frequency remains unchanged under the Lorentz transformation, i.e. $\omega_B'=\omega_B$.
In the comoving frame of the relativistic bulk motion of the pair plasma, the ratio of the gyrofrequency to the FRB angular frequency can be calculated as
\begin{equation}
\frac{\omega_B'}{\omega'}={\cal D}\frac{\omega_B}{\omega}\simeq\left\{
\begin{aligned}
&5.6\times10^{5} \ B_{\star,15}r_9^{-3}R_{\star,6}^3\nu_9^{-1}\left(\frac{\cal D}{200}\right), \\
&5.6\times10^3 \ B_{\star,15}r_9^{-3}R_{\star,6}^3\nu_9^{-1}\left(\frac{\cal D}{2}\right),
\end{aligned}
\right.
\end{equation}
where the Doppler factor is adopted as ${\cal D}\simeq 2\gamma$ for aligned propagation with $\theta_B=0$, corresponding to ${\cal D}=200$ for $\gamma=10^2$.
For oblique propagation with $\theta_B=0.1$, the Doppler factor is normalized to $2$ for $\gamma=10^2$.

At the typical radius of $r\simeq 100 R_\star$, we find $\omega_B'/\omega'\sim 6\times10^6$ for $\nu=1$ GHz, $\theta_B=0.1$ and $\gamma=100$.
In the comoving frame, the propagation angle becomes $\theta_B'\simeq0.2$.
The corresponding non-linear parameter is $a\simeq1.6\times10^5 \ L_{\rm frb,42}^{1/2}\nu_9^{-1}r_8^{-1}$, indicating that the interaction is well outside the linear scattering regime.
The scattering cross section of O-mode waves therefore needs to be evaluated numerically, yielding $\sigma'({\rm O})\simeq8.1\times10^4\sigma_{\rm T}$.
The scattering optical depth can be estimated as
\begin{equation}\label{eq:tau_O_linear}
\begin{aligned}
\tau_O&\simeq \xi n_{\rm GJ}\sigma'({\rm O})(1-\beta\cos\theta_B)r\\
&\simeq9.5\times10^{-7} \ \xi_2 r_8^{-2}B_{\star,15}R_{\star,6}^3P^{-1}\left(\frac{\sigma'}{8.1\times10^4\sigma_{\rm T}}\right)\ll1,
\end{aligned}
\end{equation}
where the multiplicity factor $\xi=10^2$ is adopted as a typical value.
We therefore conclude that relativistic particle motion does not lead to significant scattering losses for O-mode FRB waves in the linear magnetospheric region.

\section{Numerical Results}\label{sec:Numerical results}

In this section, we present our numerical results obtained by solving the test charged particle equation of motion in a static background magnetic field and in the presence of large-amplitude linearly polarized X-mode, O-mode, and mixed-mode waves.
We compute the scattering cross section as a function of the ratio between the cyclotron frequency and the angular frequency of FRBs, $\omega_B'/\omega'$, and normalize it by $a^2\sigma_{\rm T}$. 
The initial phase of the wave is set to $\psi_0=0$. 
The time resolution is chosen to 
resolve the particle gyration motion, particularly in the regime of large $\omega_B'/\omega'$.
For a given set of parameters, the simulations must therefore be evolved for a sufficiently long time to obtain a reliable time-averaged cross section $\bar{\sigma}'$.
In our numerical results, our $\bar\sigma'$ values are typically derived for a time duration $t \sim 10^6 \nu_{\rm FRB}^{-1} \sim 1 \ \nu_{\rm GHz}^{-1} \ \rm ms$. 
This millisecond timescale is consistent with the typical FRB duration.
The calculated cross section can be regarded as a conservative upper limit because radiation reaction is not included and particles are expected to rapidly lose energy through radiative cooling rather than continuously gaining energy from the incoming FRB waves.

The propagation angle $\theta_B'$ in the comoving frame is related to the lab frame angle through $\sin\theta_B'={\cal D}\sin\theta_B$, where ${\cal D}={1}/{[\gamma(1-\beta\cos\theta_B)}]$ is the Doppler factor.
When FRB waves propagate through the open field line region, the angle $\theta_B$ varies from zero up to $\lesssim 10^\circ$ in a dipolar magnetic geometry \citep{QKZ}. 
In this region, particles stream relativistically along the magnetic field. As a result, $\theta_B'$ can be the same order as $\theta_B$ and even smaller as long as $\gamma\gtrsim 10^2$.
By contrast, in the closed field line region particles are non-relativistic. 
In this case, the transformation is negligible and we have $\theta_B'\simeq \theta_B\lesssim90^\circ$.
We note that there exists a special propagation angle for which $\theta_B = 1/\gamma$ in the lab frame corresponds to $\theta_B' = \pi/2$ in the comoving frame. 
Although this configuration formally maps to perpendicular propagation in the comoving frame, it represents a finely tuned condition. 
As FRB waves propagate through the magnetosphere, the variation of $\theta_B$ with radius is continuous, and the wave crosses this specific angle over an extremely short spatial and temporal interval. 
Consequently, the system cannot remain at $\theta_B = 1/\gamma$ for a duration comparable to a typical FRB timescale. 
We therefore do not expect this special configuration to play a significant role in the overall wave -- particle interaction.

In the following, we investigate the evolution of the particle Lorentz factor and the scattering cross section for these representative values of $\theta_B'$ in the comoving frame.

\subsection{Linearly Polarized Waves}

\begin{figure*}
\begin{center}
\setlength{\tabcolsep}{-8pt}
\begin{tabular}{lll}
\resizebox{67mm}{!}{\includegraphics[]{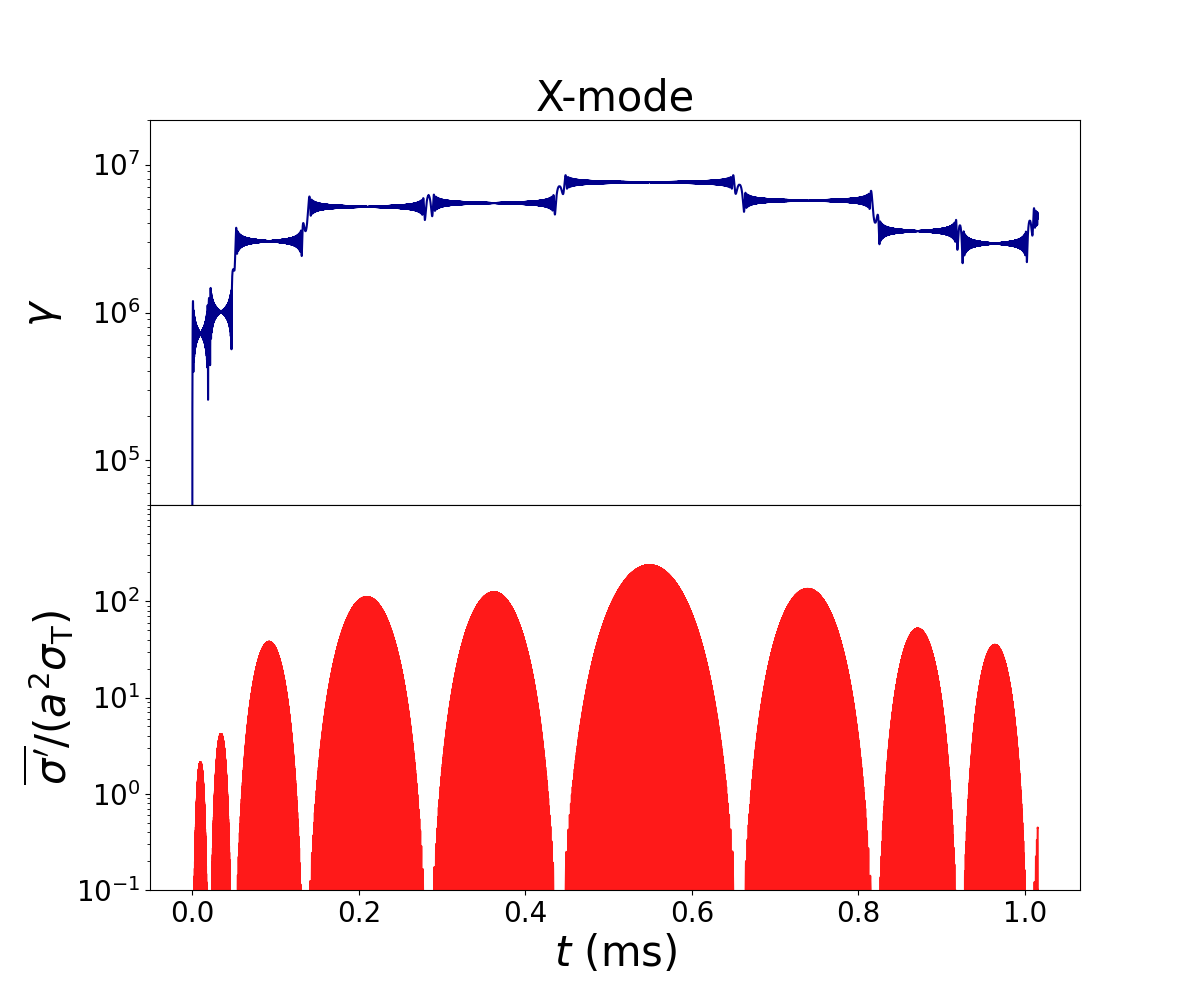}}&
\resizebox{67mm}{!}{\includegraphics[]{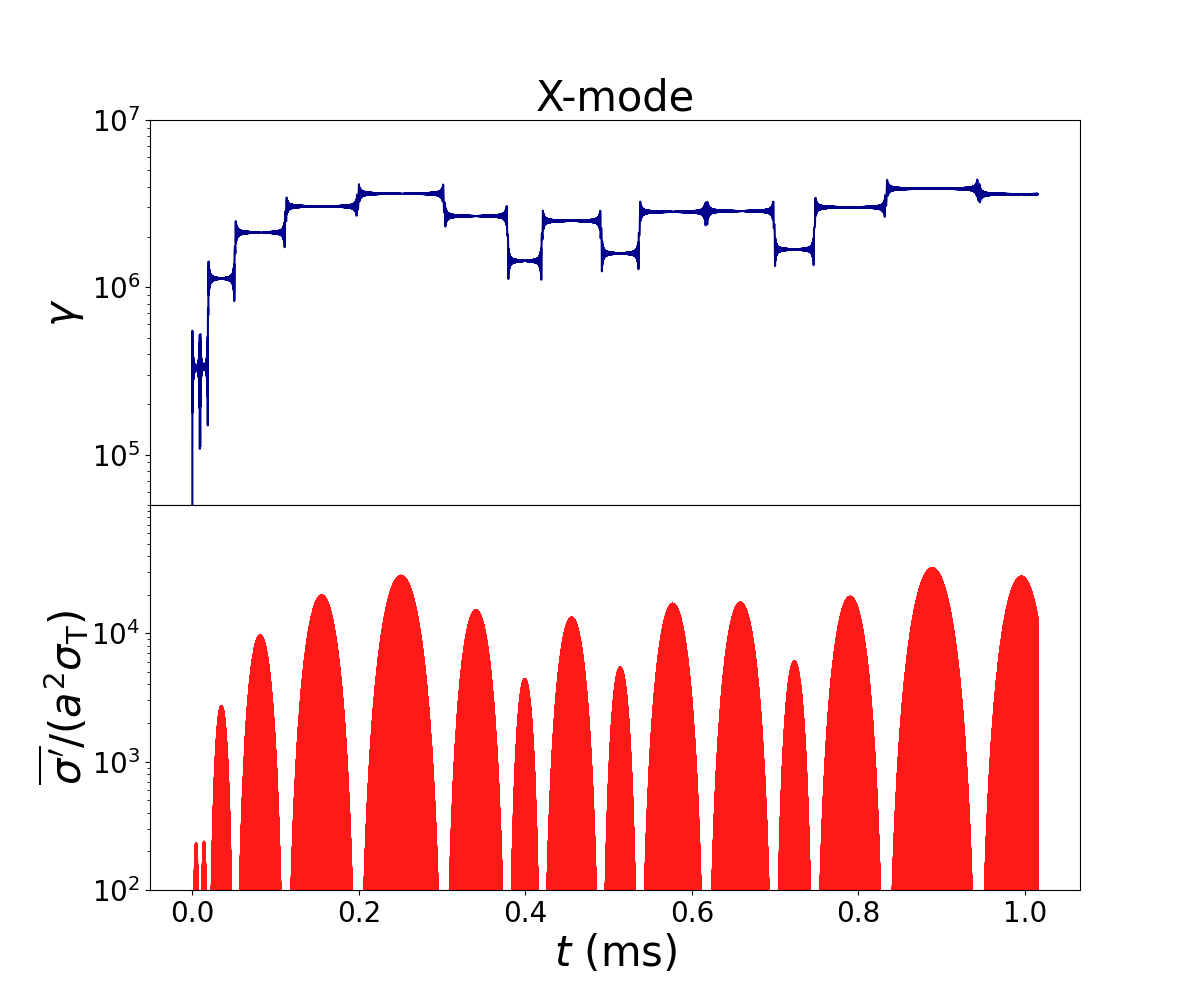}}&
\resizebox{67mm}{!}{\includegraphics[]{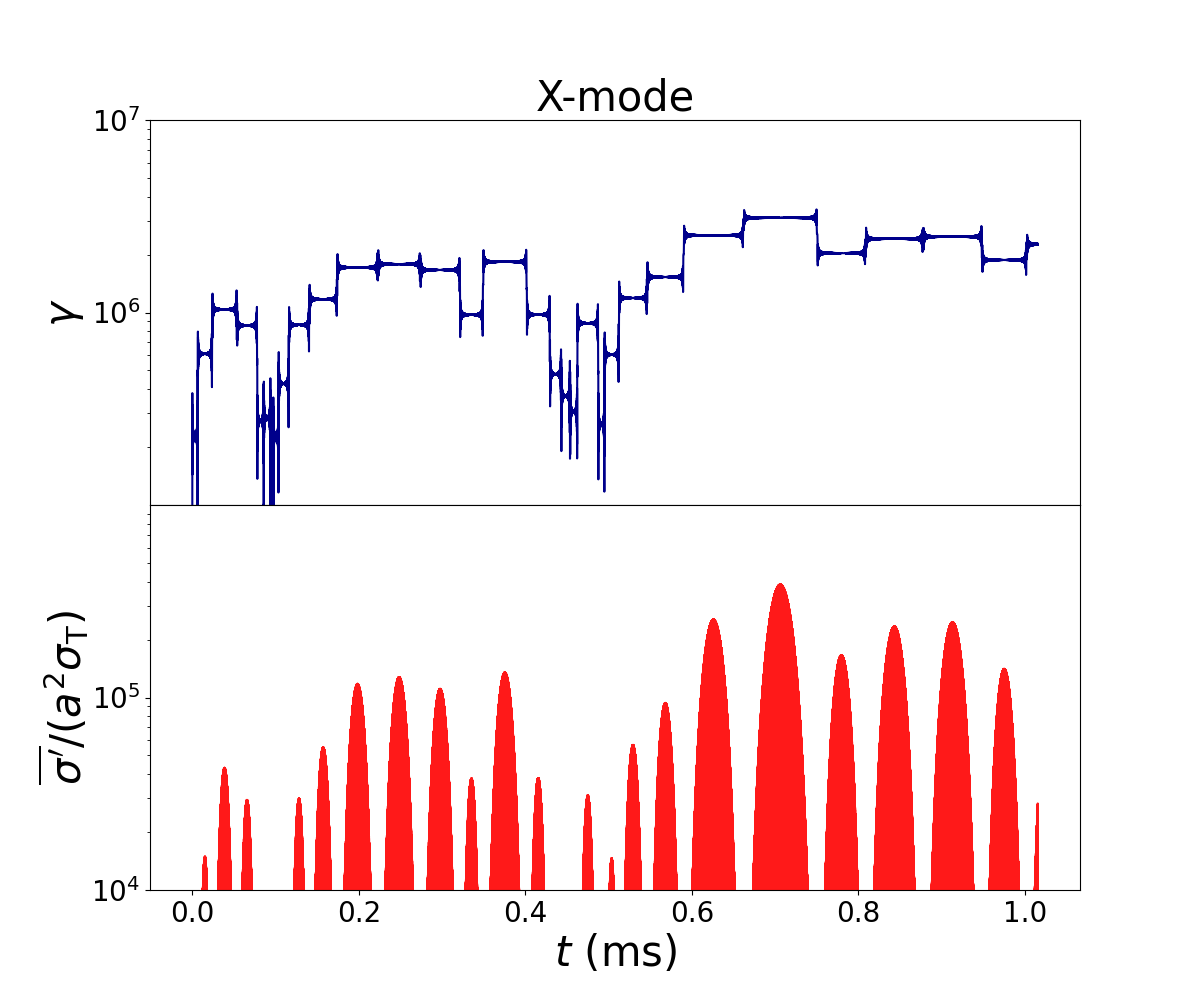}}\\
\resizebox{67mm}{!}{\includegraphics[]{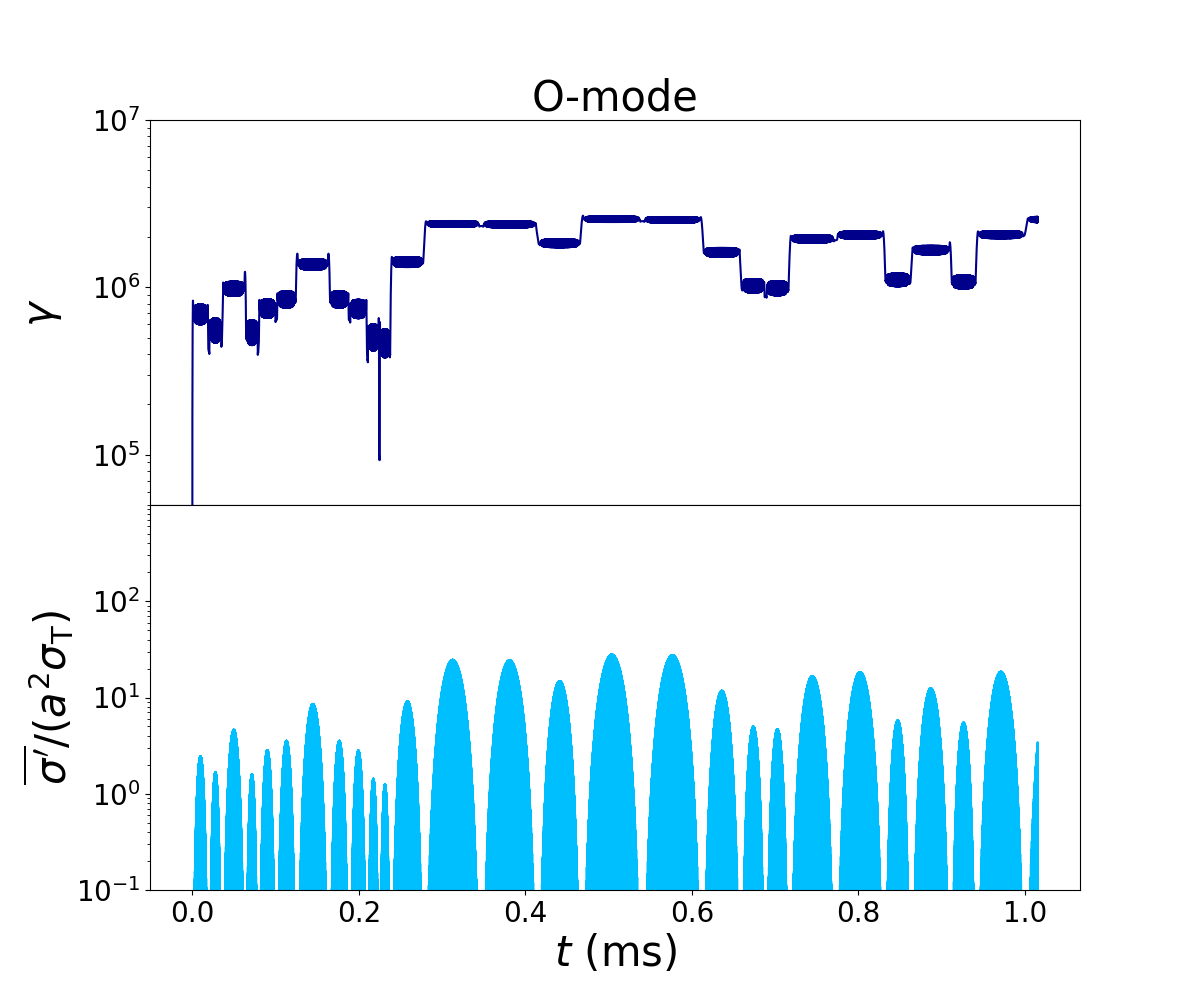}}&
\resizebox{67mm}{!}{\includegraphics[]{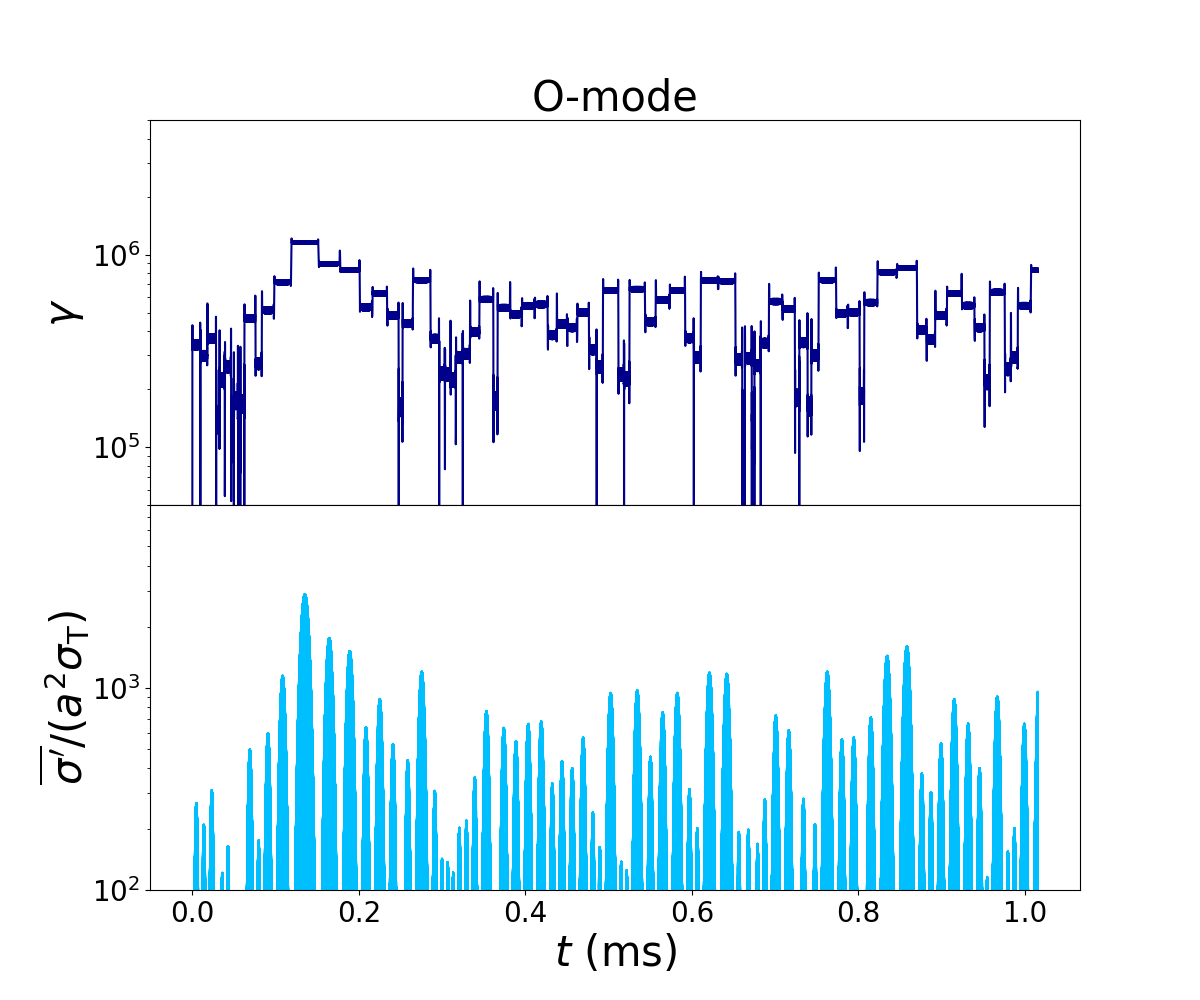}}&
\resizebox{67mm}{!}{\includegraphics[]{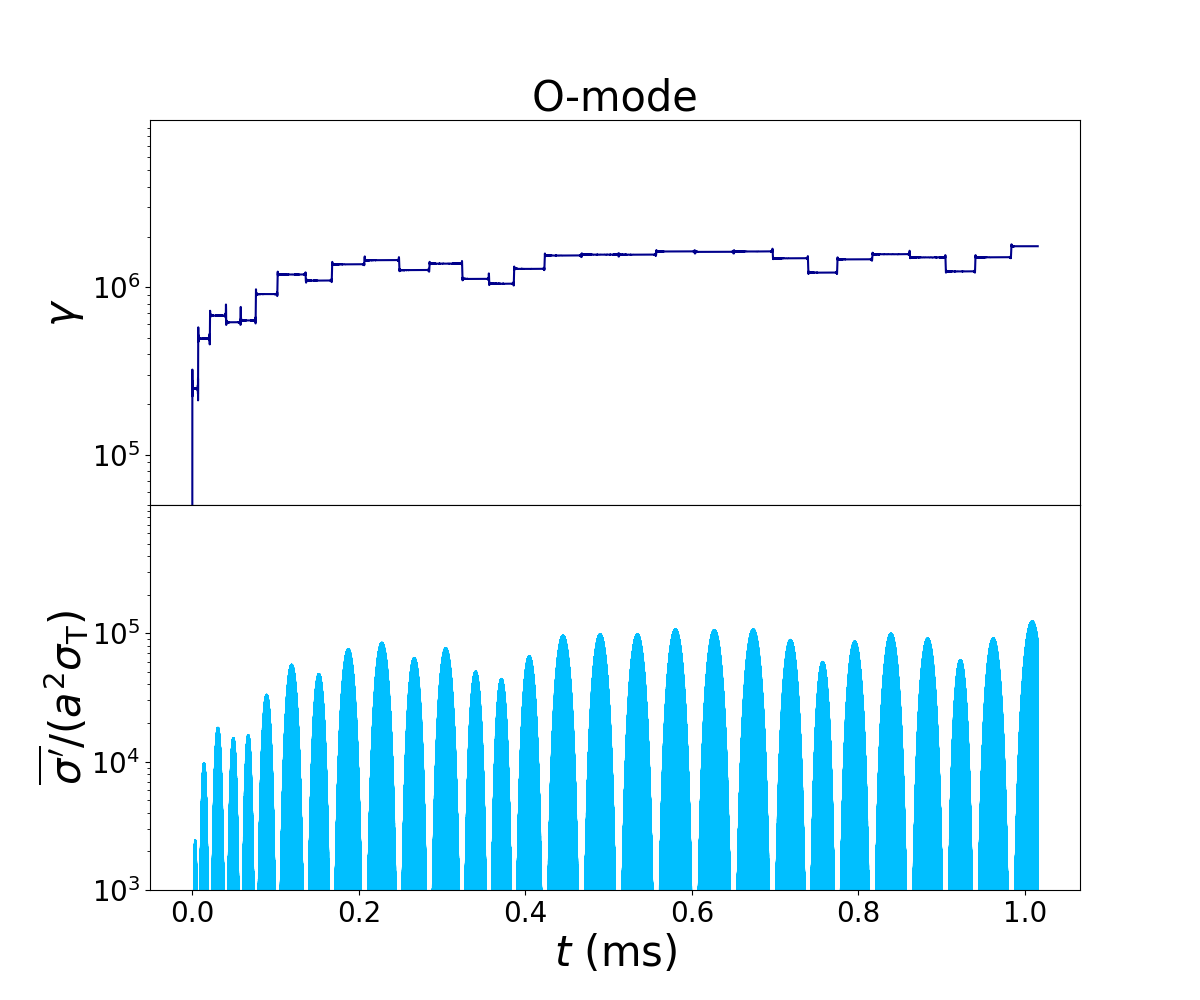}}\\
\resizebox{67mm}{!}{\includegraphics[]{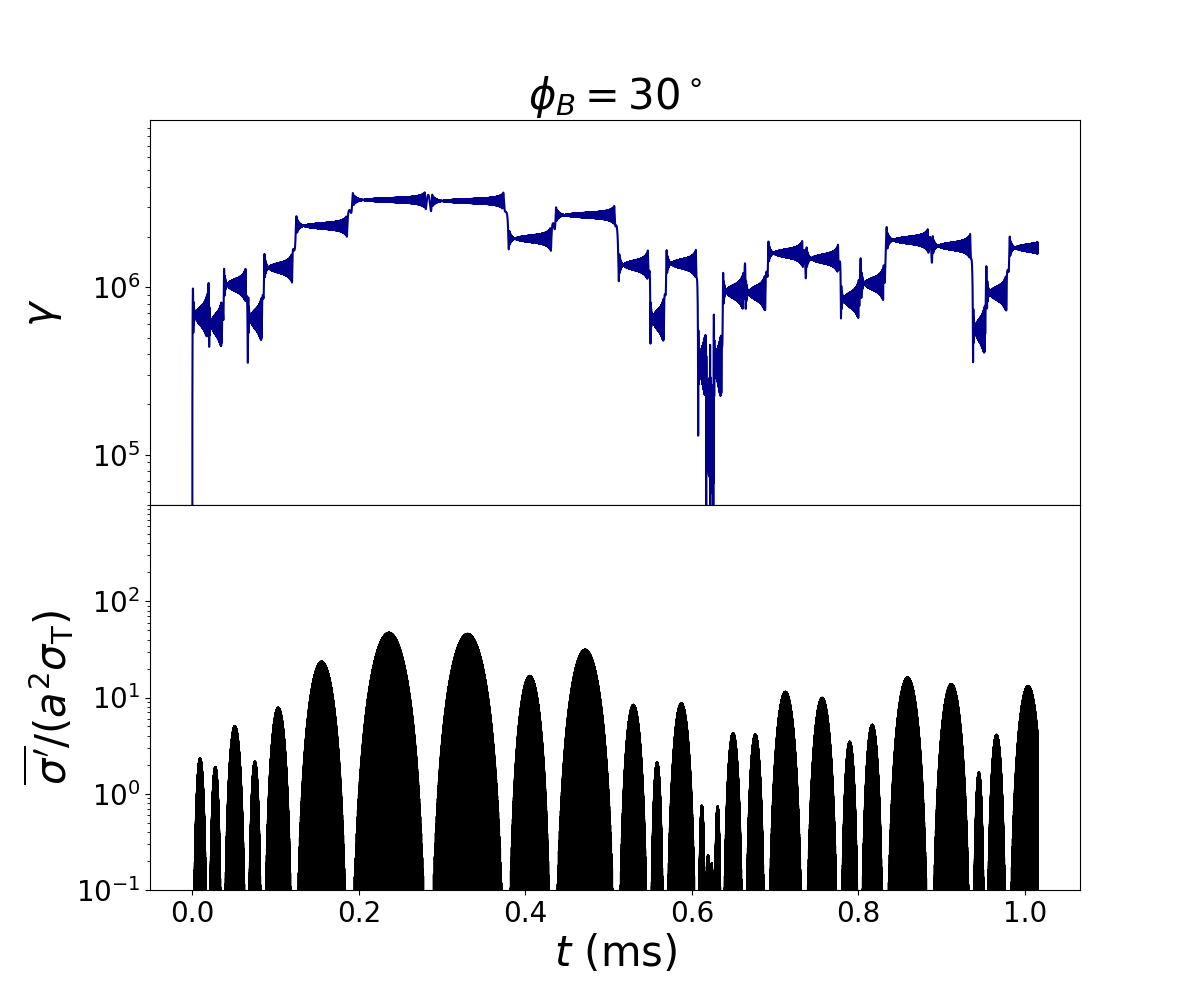}}&
\resizebox{67mm}{!}{\includegraphics[]{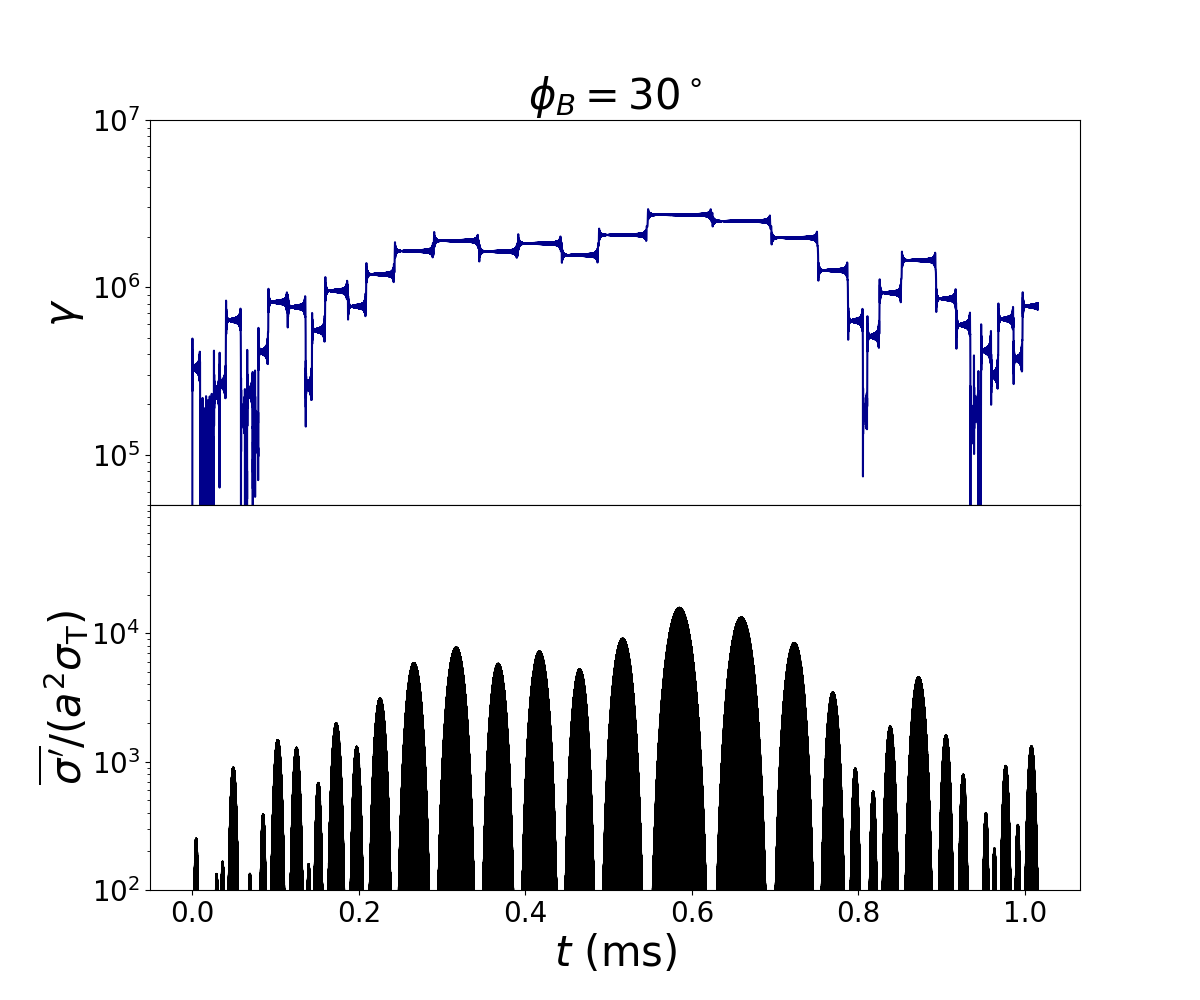}}&
\resizebox{67mm}{!}{\includegraphics[]{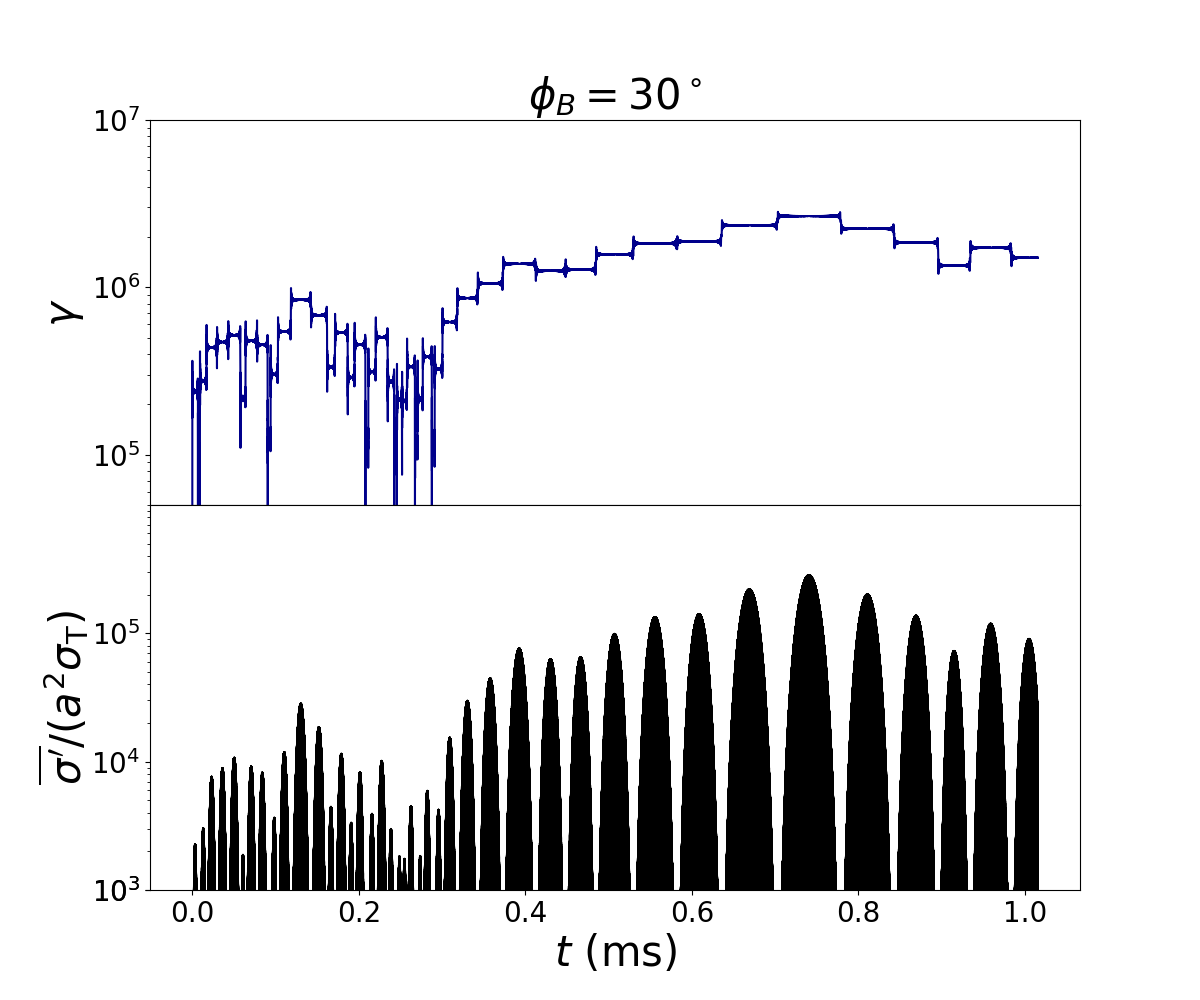}}\\
\end{tabular}
\caption{The electron Lorentz factor (upper subpanel) and normalized scattering cross section (lower subpanel) as a function of time for different $\theta_{B}'=0.1, 0.5, \pi/2$ for left, central and right panels, respectively. 
The non-linear parameter $a=10^4$ and $\omega_B'/\omega'=35$ are adopted for all panels.}
\label{fig:mode} 
\end{center}
\end{figure*}

\begin{figure*}
\begin{center}
\hspace*{-6mm}
\begin{tabular}{ll}
\resizebox{86mm}{!}{\includegraphics[]{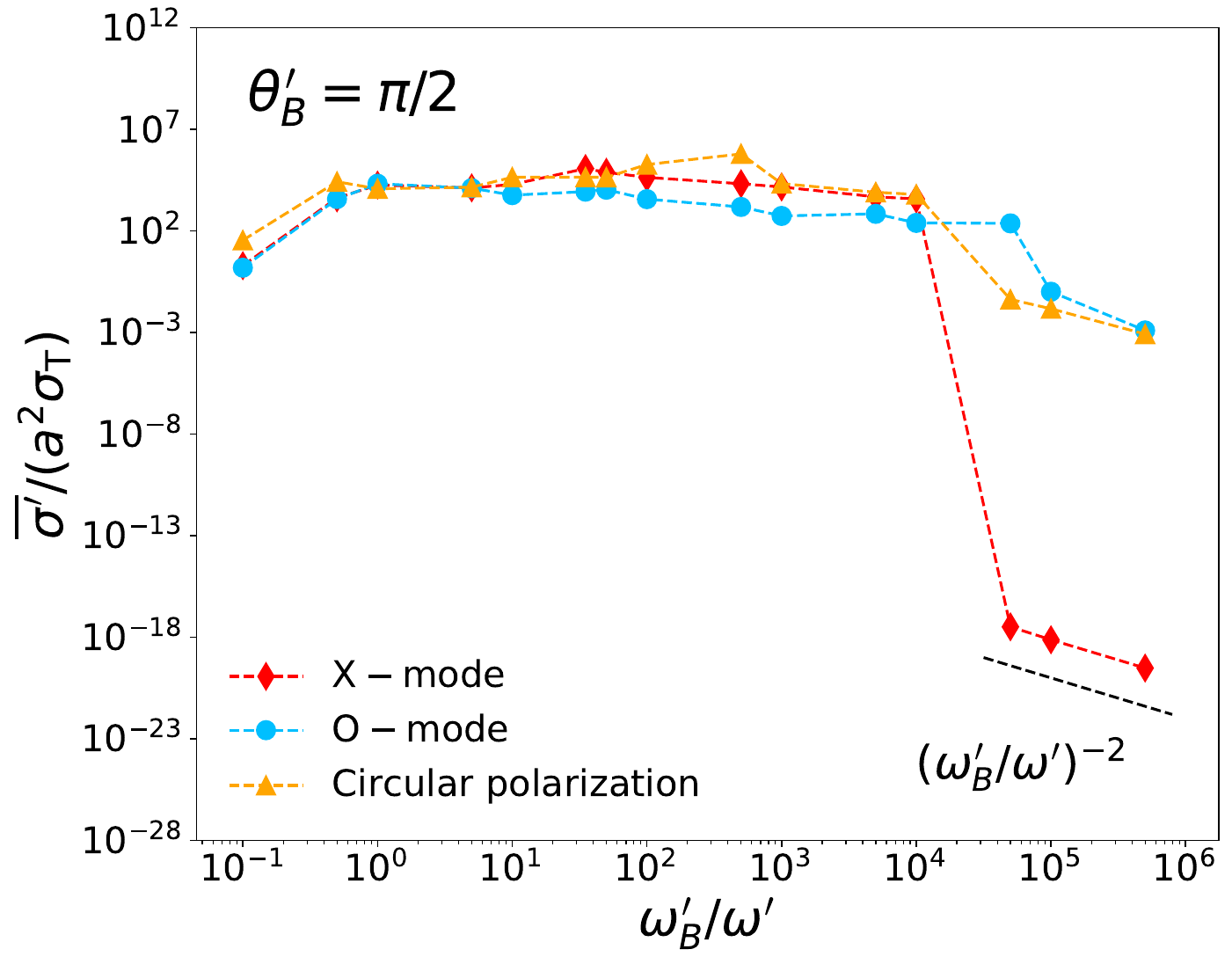}}&
\resizebox{86mm}{!}{\includegraphics[]{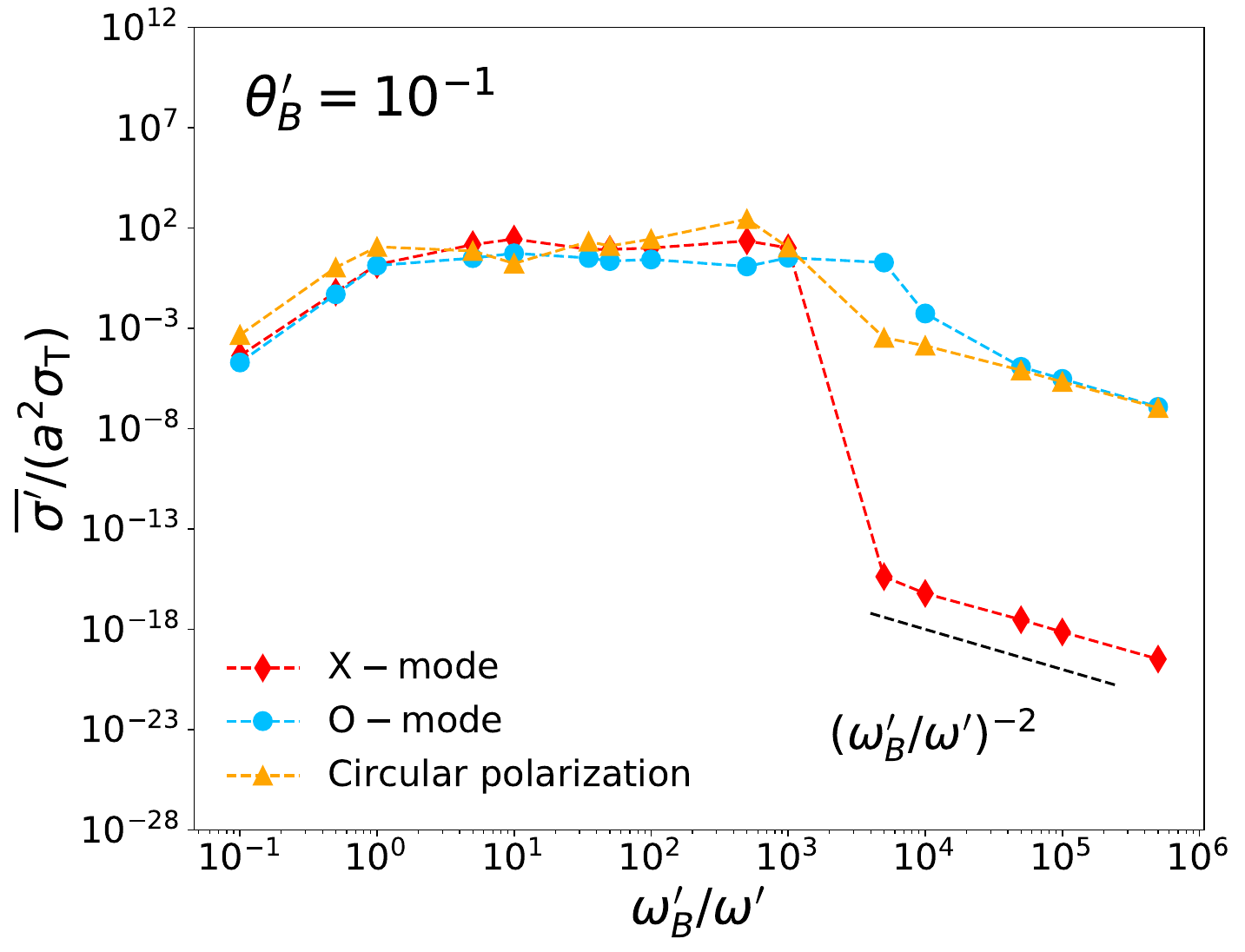}}\\
\resizebox{86mm}{!}{\includegraphics[]{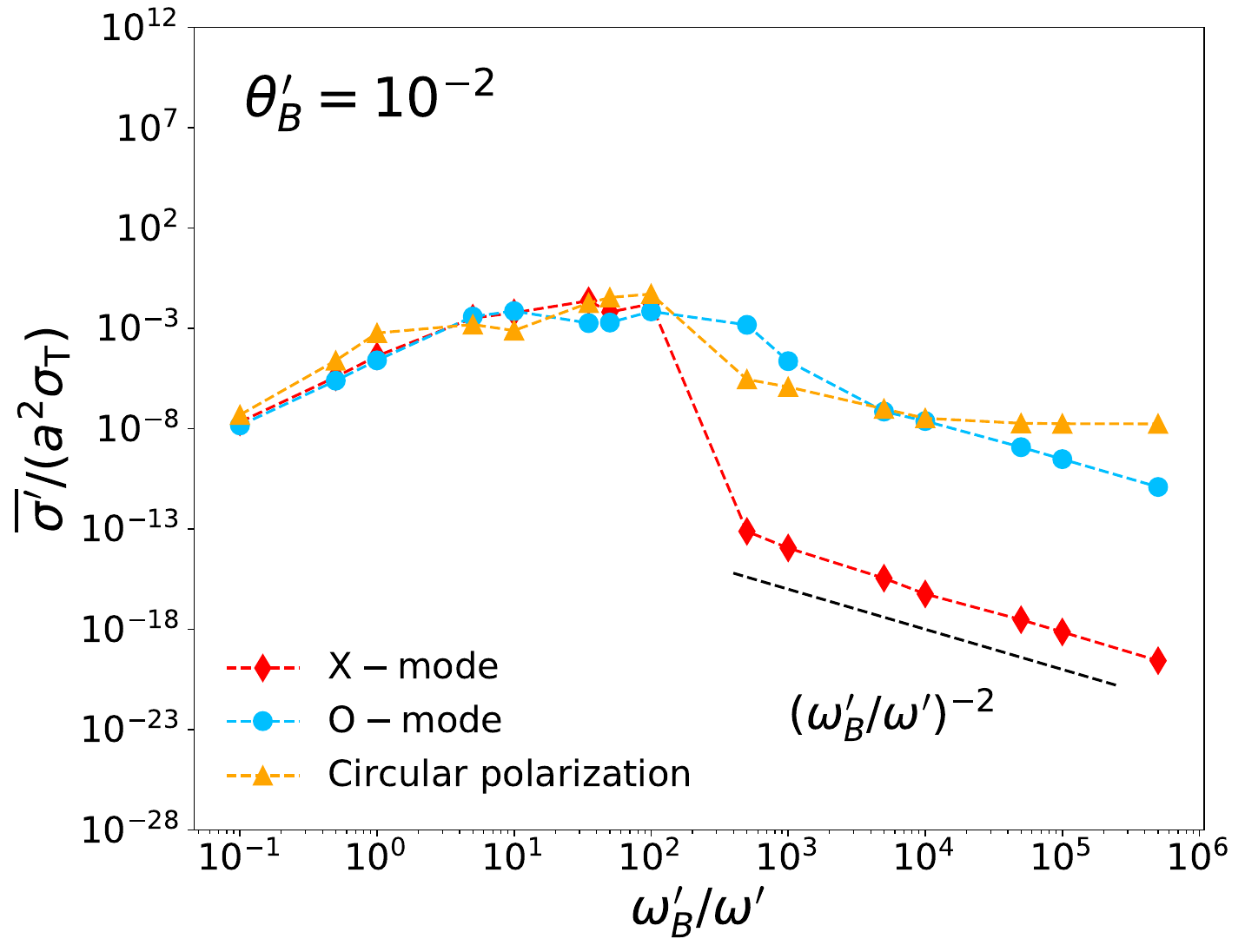}}&
\resizebox{86mm}{!}{\includegraphics[]{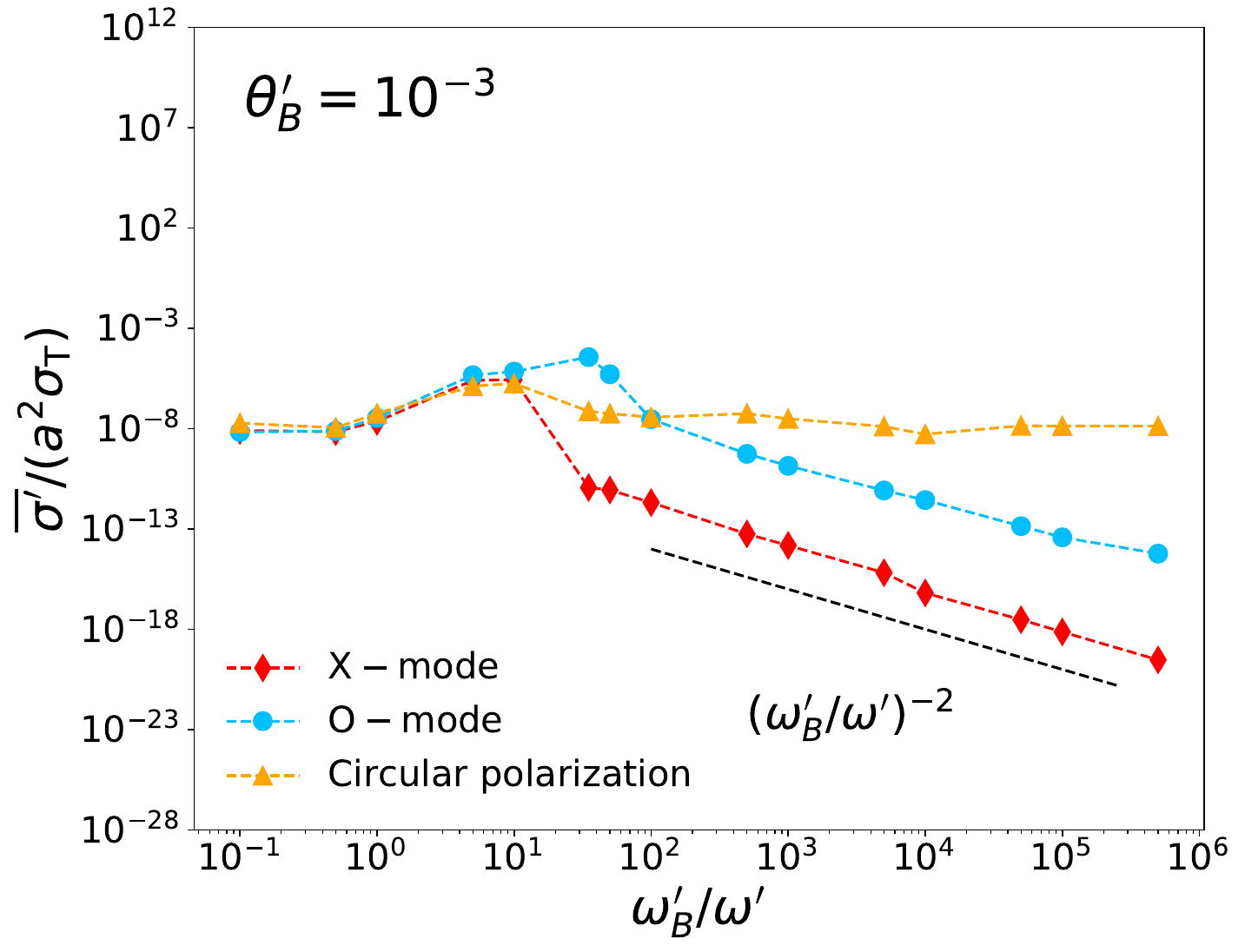}}
\end{tabular}
\caption{The normalized scattering cross section $\overline{\sigma'}/(a^2\sigma_{\rm T})$ as a function of $\omega'_B/\omega'$ for different propagation angles $\theta'_B$.
The four panels correspond to $\theta'_B = \pi/2, 10^{-1}, 10^{-2}, 10^{-3}$ (from top left to bottom right).
Red, blue, and orange curves denote X-mode, O-mode, and circularly polarized waves, respectively.
The black dotted lines indicate the characteristic scalings in different regimes: $\sigma' \propto (\omega'_B/\omega')^{3}$ in the nonlinear regime and $\sigma'\propto (\omega_B'/\omega')^{-2}$ in the linear regime.
Wave strengths $a_{\rm LP}=10^4$ and $a_{\rm CP}=a_{\rm LP}/\sqrt{2}$ are adopted for linearly and circularly polarized waves, respectively.
}
\label{fig:sigma_omega_B}
\end{center}
\end{figure*}

\begin{figure*}
\begin{center}
\begin{tabular}{ll}
\resizebox{85mm}{!}{\includegraphics[]{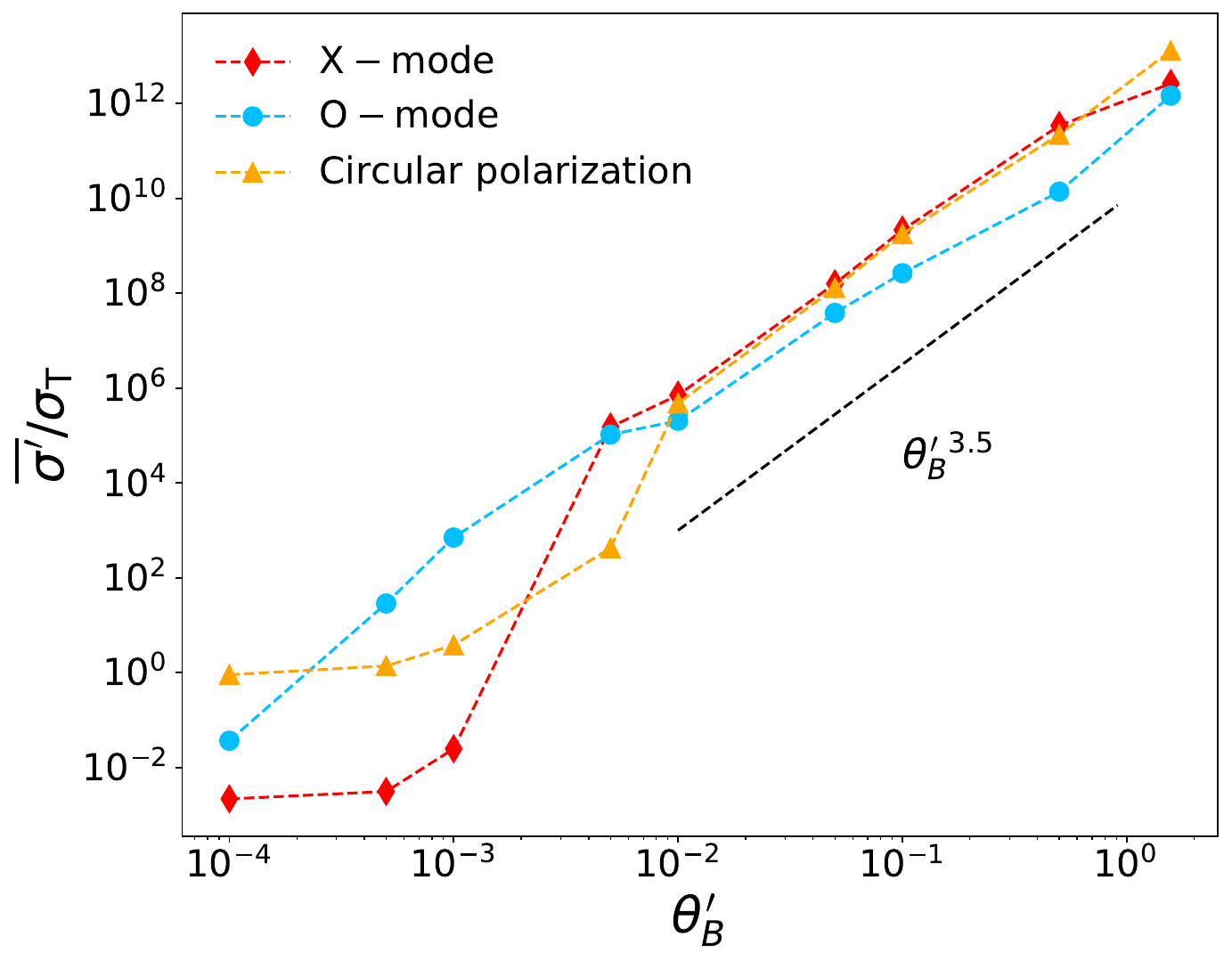}}&
\resizebox{85mm}{!}{\includegraphics[]{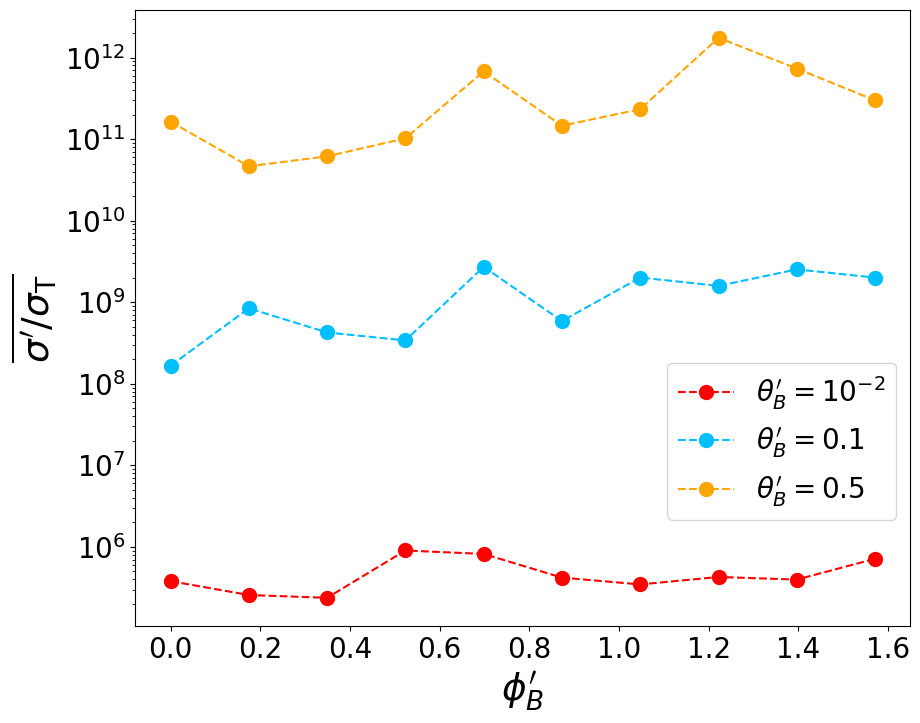}}
\end{tabular}
\caption{The normalized cross section $\overline{\sigma'}/\sigma_{\rm T}$ as a function of $\theta_B'$ (left panel) and $\phi_B'$ (right panel). Following parameters are adopted: $a_{\rm LP}=10^4$, $a_{\rm CP}=a_{\rm LP}/\sqrt{2}$, and $\omega_B'/\omega'=35$.}
\label{fig:sigma}
\end{center}
\end{figure*}

For FRBs with $a=10^4$ (Equation~(\ref{eq:typical_a_parameter})) propagating in the outer magnetosphere,
the numerical results for the particle Lorentz factor $\gamma$ and the normalized cross section $\overline{\sigma}'/(a^2\sigma_{\rm T})$ are shown in Figure~\ref{fig:mode}.
The upper, middle, and lower panels correspond to X-mode, O-mode, and mixed-mode waves, respectively, plotted as functions of $\omega_B'/\omega'$.
Both the Lorentz factor and the normalized cross section exhibit strong temporal variations.
From left to right, the three columns correspond to $\theta_B'=0.1, 0.5$, and $\pi/2$, respectively.
A clear trend emerges that the scattering cross section is strongly suppressed as $\theta_B'$ decreases for all three initial polarization states.

In the upper left panel of Figure~\ref{fig:sigma_omega_B} ($\theta_B'=10^{-1}$), the normalized cross sections of both X-mode and O-mode waves initially increase with $\omega_B'/\omega'$ at small values and enter a plateau with a maximum value of order $\sim 10$. 
Beyond a characteristic transition point, the cross sections drop rapidly to values much smaller than unity.
The decrease is more pronounced for X-mode waves than for O-mode waves.
At sufficiently large $\omega_B'/\omega'$, the cross sections of X-mode waves converge and decrease with a common slope of $\sim-2$. 
The O-mode cross section initially follows the same trend as the X-mode but subsequently approaches a plateau with $\sigma'\sim \sigma_{\rm T}\sin^2\theta_B'$. 
This behavior corresponds to the linear scattering regime and is consistent with the results shown in Figure~\ref{fig:sigma_Omode}.
The O-mode retains an unscreened longitudinal electric field component and thus has a larger cross section at larger $\omega_B'/\omega'$ when $\theta_B'>0$, while the X-mode becomes increasingly suppressed as the particle motion is constrained by the strong magnetic field.
For X-mode waves, the termination of nonlinear scattering occurs when the wave induced motion becomes subdominant to the cyclotron motion. 
This condition can be expressed as
\begin{equation}
a'\sin\theta_B'\lesssim\frac{\omega_B'}{\omega'},
\end{equation}
below which nonlinear scattering is no longer effective and the interaction enters the linear scattering regime \citep{QKZ}.
For O-mode waves, the termination of nonlinear scattering is delayed to larger values of $\omega_B'/\omega'$, lagging behind the X-mode transition by approximately one order of magnitude.
This delay is captured and the termination of nonlinear scattering condition can be written as 
\begin{equation}
a'\sin\theta_B'=\frac{\omega_B'}{\zeta_O\omega'}, \  \zeta_O\sim10,
\end{equation}
which indicates that O-mode waves require a stronger magnetic dominance, or equivalently a larger $\omega_B'/\omega'$, to suppress nonlinear scattering.

The appearance of the factor $\sin\theta_B'$ in the non-linear condition can be understood from a Lorentz-frame argument.
For an obliquely propagating wave in the particle rest frame, one can always perform a Lorentz boost along the wave propagation direction such that the magnetic field becomes perpendicular to the wave vector in the new frame.
This corresponds to a Lorentz factor $\gamma=1/\sin\theta_B'$.
In this frame, while the wave frequency is Doppler shifted, the magnetization parameter entering the particle dynamics is enhanced by the geometric factor $1/\sin\theta_B'$. 
As a result, the scattering process is governed by the combination $\gamma \omega_B'/\omega'$.
Consequently, for sufficiently small $\theta_B'$, the magnetic confinement becomes strong and suppresses the scattering cross section.
O-mode waves exhibit a similar behavior, but the transition occurs at systematically larger $\omega_B'/\omega'$, reflecting the presence of an unscreened longitudinal electric field component.
Relativistic pairs stream predominantly along the background magnetic field.
For Lorentz boosts parallel to the field, the cyclotron frequency is invariant, $\omega_B'=\omega_B$.
In the comoving frame of the relativistic plasma, the ratio between the cyclotron and wave frequencies and the Doppler factor are
\begin{equation}
\frac{\omega_B'}{\omega'}={\cal D}\frac{\omega_B}{\omega}, \ {\cal D}=\frac{1}{\gamma(1-\beta\cos\theta_B)}.
\end{equation}
For quasi-parallel propagation with $\gamma\theta_B\ll1$, ${\cal D}\approx 2\gamma$, and $\omega_B'/\omega'$ can be orders of magnitude larger than unity even for moderate Lorentz factors.
The resulting cross section can remain in the linear regime despite $E_w>B$ in the lab frame.
This demonstrates that for relativistic particles the criterion $E_w > B$ is not sufficient to ensure nonlinear scattering.
Even when the wave field exceeds the background field in the lab frame, scattering can remain weak if the propagation is quasi-parallel and the particle motion is relativistic.

In the upper right panel of Figure~\ref{fig:sigma_omega_B} ($\theta_B'=10^{-1}$), the normalized cross sections are systematically smaller than those in the upper left panel. 
They show a similar trend, increasing initially with $\omega_B'/\omega'$ and then decreasing sharply beyond the transition point. 
In the lower left and lower right panels ($\theta_B'=10^{-2}$ and $\theta_B'=10^{-3}$), the cross sections are further reduced. 
They remain nearly flat as a function of $\omega_B'/\omega'$ before undergoing a rapid decline at the transition. 
When $\theta_B'=10^{-3}$, the cross sections of X-mode and O-mode become nearly indistinguishable and reach comparable values. 
This convergence is expected because the wave vector is almost perfectly aligned with the background magnetic field, which eliminates the polarization dependent dynamics.

For a fixed $\omega_B'/\omega'=35$, we present $\overline{\sigma'}/\sigma_{\rm T}$ as a function of $\theta_B'$ in the left panel of Figure~\ref{fig:sigma}. 
One can see that the transition to the suppressed scattering regime for X-mode occurs at
$a\sin\theta_B'\lesssim\omega_B'/\omega'$, which defines the characteristic boundary between strong and weak scattering.
The cross section remains nearly constant at small $\theta_B'$ and increases rapidly once the transition condition is crossed. 
The cross sections of X-mode, O-mode, and circularly polarized waves follow an empirical scaling
$\sigma' \propto \theta_B'^{3.5}$ at larger $\theta_B'$.
For X-mode, we have
\begin{equation}\label{eq:sigma_X}
\sigma'({\rm X}) \propto \theta_B'^{\ 3.5}, \ \theta_B'>\frac{\omega_B'}{a\omega'},
\end{equation}
and
\begin{equation}\label{eq:sigma_O}
\sigma'({\rm O}) \propto \theta_B'^{\ 3.5}
\end{equation}
for O-mode.
O-mode cross sections remain larger than those of X-mode at small but finite $\theta_B'$, while both cross sections converge and become nearly indistinguishable as $\theta_B'\rightarrow0$, indicating that polarization-dependent dynamics are eliminated when the wave propagates nearly along the magnetic field.
These results demonstrate a strong sensitivity of the scattering cross section to the propagation angle. Even a modest change in $\theta_B'$, for example by a factor of two, can lead to an order of magnitude variation in the cross section.

We also present the dependence of $\overline{\sigma'}/\sigma_{\rm T}$ on the azimuthal angle of the background magnetic field $\phi_B'$ in the right panel of Figure~\ref{fig:sigma}.
The limits $\phi_B'=0$ and $\phi_B'=\pi/2$ correspond to pure O-mode and pure X-mode waves, respectively.
Intermediate values $0<\phi_B'<\pi/2$ represent linearly polarized mixed modes.
We find that the scattering cross section varies by roughly an order of magnitude across this range of $\phi_B'$. 
This behavior reflects the continuous change in the relative contributions of the electric field components parallel and perpendicular to the background magnetic field.

\subsection{Circularly Polarized Waves}

It should be pointed that fixing the non-linear parameter does not imply a fixed wave intensity when comparing different polarization states.
For a linearly polarized wave with electric field $\vec{E}_{\rm LP}=E_0\sin\psi\hat{x}$, the field amplitude oscillates in time and the cycle-averaged squared field is $\langle E_{\rm LP}^2\rangle=E_0^2/2$, corresponding to an average Poynting flux $\langle S\rangle_{\rm LP}=(c/8\pi)E_0^2$. 
In contrast, a circularly polarized wave with $\vec{E}_{\rm CP}=E_0(\sin\psi\hat{x}+\cos\psi\hat{y})$ has a constant field magnitude $|\vec{E}_{\rm CP}|=E_0$ at all phases, yielding $\langle E_{\rm CP}^2\rangle=E_0^2$ and thus $\langle S\rangle_{\rm CP}=(c/4\pi)E_0^2$.
As a result, at the same value of $a$-parameter, a circularly polarized wave carries a factor of two larger Poynting flux than a linearly polarized wave. 
Thus we rescale the circularly polarized field amplitude as $E_0\rightarrow E_0/\sqrt{2}$ and $a_{\rm CP}=a_{\rm LP}/\sqrt{2}$.

\begin{figure*}
\begin{center}
\setlength{\tabcolsep}{-8pt}
\begin{tabular}{lll}
\resizebox{67mm}{!}{\includegraphics[]{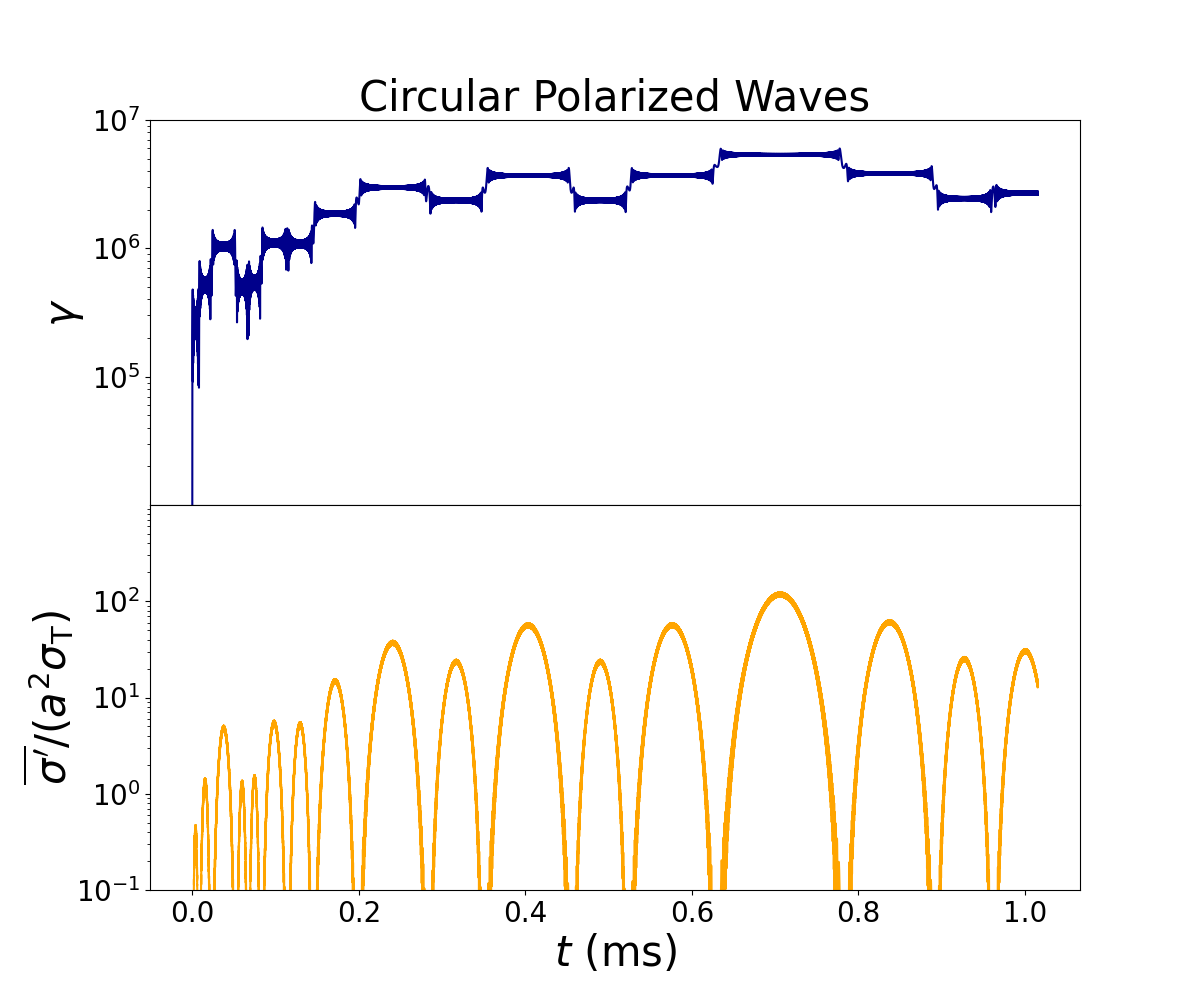}}&
\resizebox{67mm}{!}{\includegraphics[]{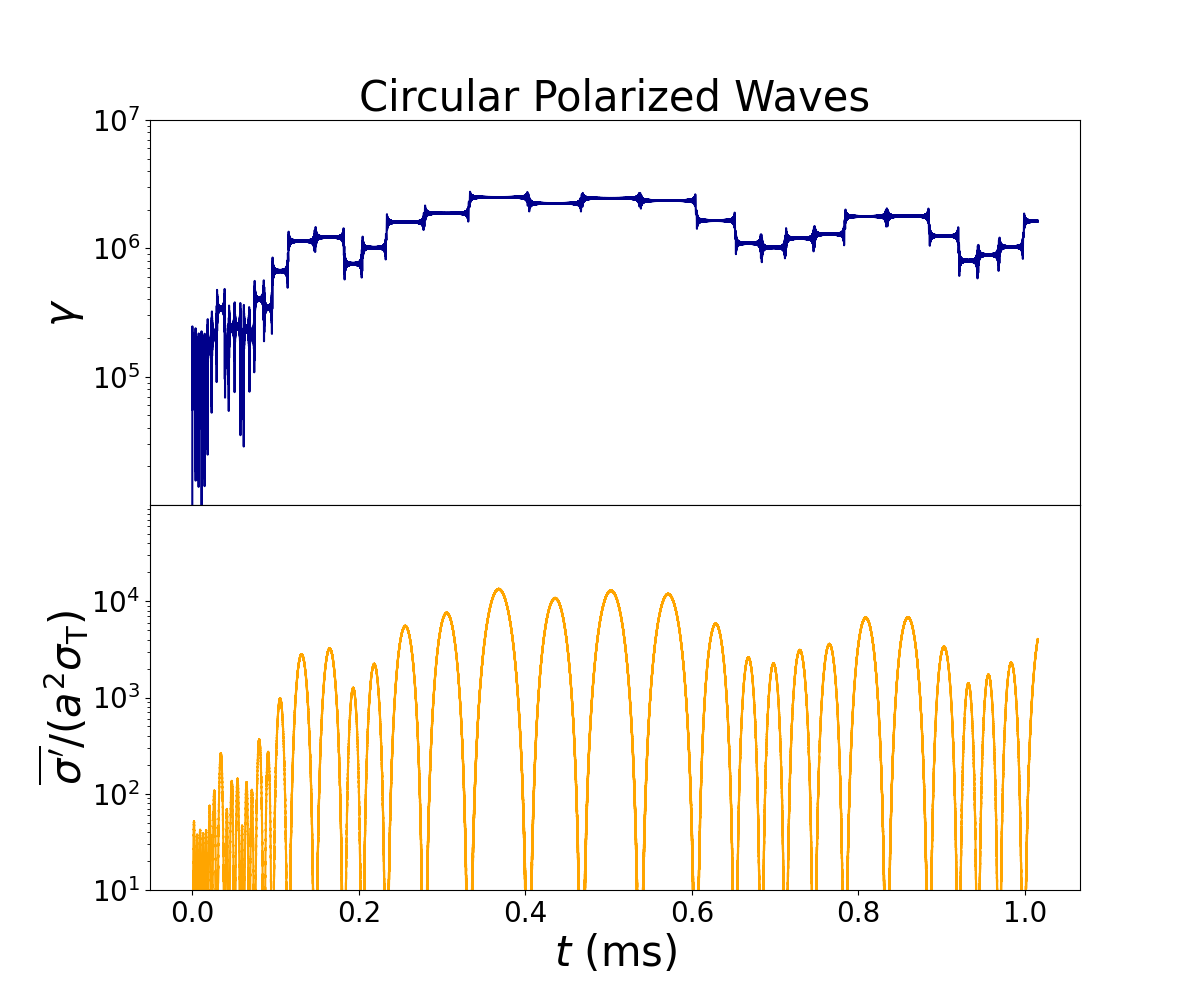}}&
\resizebox{67mm}{!}{\includegraphics[]{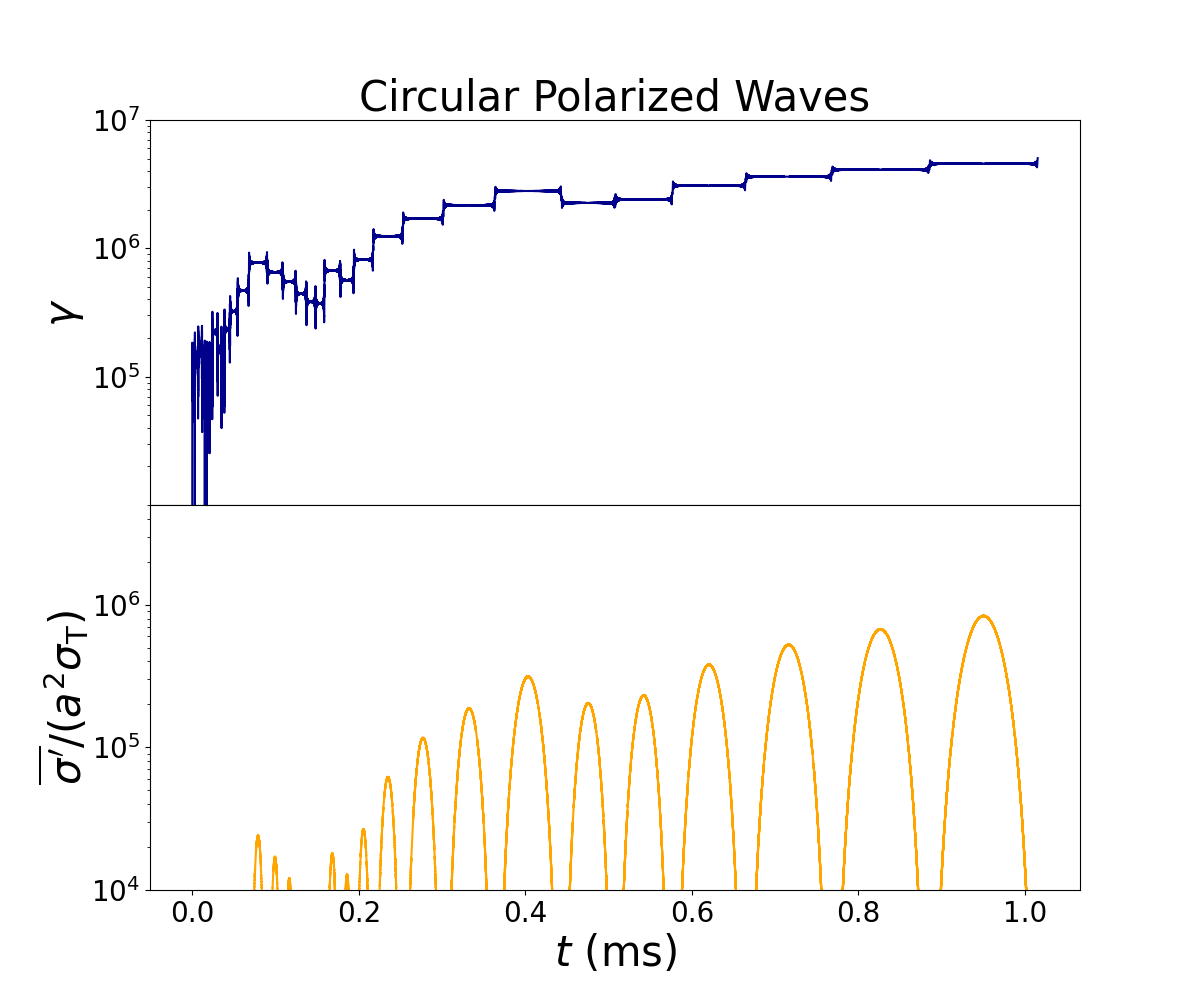}}\\
\end{tabular}
\caption{Same as Figure~\ref{fig:mode} but for different $\theta_{B}'=0.1, 0.5, \pi/2$ for left, central and right panels, respectively, but for 100\% circularly polarized waves with $a_{\rm CP}=a_{\rm LP}/\sqrt{2}$ and $a_{\rm LP}=10^4$.}
\label{fig:cirmode} 
\end{center}
\end{figure*}

In the left panel of Figure~\ref{fig:sigma}, we present the scattering cross sections of fully circularly polarized waves as a function of the propagation angle $\theta_B'$.
The cross sections are generally larger than those of both the X- and O-modes over the entire range of $\theta_B'$.
Similar to the O-mode behavior, the circularly polarized cross sections follow an empirical scaling $\propto \theta_B'^{\ 3.5}$ across the full angular range considered.
As in Figure~\ref{fig:mode}, we further show in Figure~\ref{fig:cirmode} the temporal evolution of the particle Lorentz factor and the normalized scattering cross section.

\subsection{Scattering Optical Depth}

We now estimate the scattering optical depth in the presence of relativistic streaming plasma.
For particles moving along the background magnetic field $B$ with the Lorentz factor $\gamma$, the optical depth in the laboratory frame can be approximated as $\tau\approx n \sigma'(1-\beta\cos\theta_B)l$,
where plasma number density is $n=\xi n_{\rm GJ}$, $\xi$ is the pair multiplicity, $n_{\rm GJ}$ is the Goldreich--Julian density \citep{Goldreich&Julian1969} and $l$ is the electromagnetic wave propagation distance.
In the quasi-parallel propagation limit $\theta_B\ll1$ and $\gamma\gg1$, the optical depth scales as
\begin{equation}
\tau_{\rm X}\approx \xi n_{\rm GJ}\sigma'(\gamma^{-2}+\beta\theta_B^2)l\sim \sigma'\gamma^{-2}l
\end{equation}
for X-mode and
\begin{equation}
\tau_{\rm O}\approx \xi n_{\rm GJ}\theta_B'^{3.5}\gamma^{-2}l.
\end{equation}
for O-mode and circularly polarized waves. 
These scalings illustrate that even modest alignment between the wave vector and the magnetic field, combined with relativistic streaming of the plasma, can efficiently suppress the scattering optical depth.

\begin{figure*}
\begin{center}
\setlength{\tabcolsep}{-0pt}
\begin{tabular}{lll}
\resizebox{60mm}{!}{\includegraphics[]{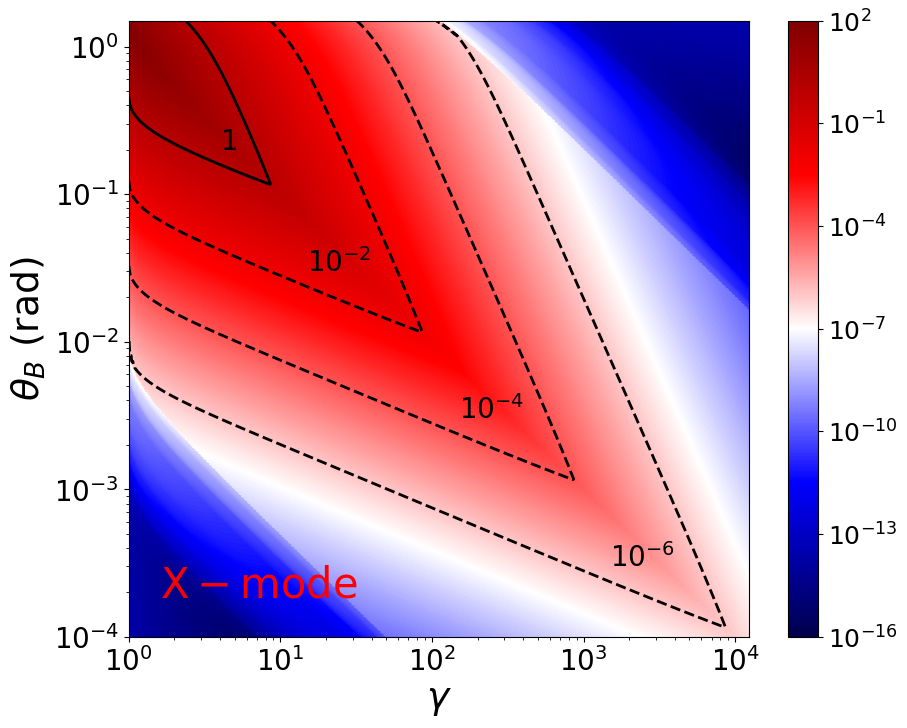}}&
\resizebox{60mm}{!}{\includegraphics[]{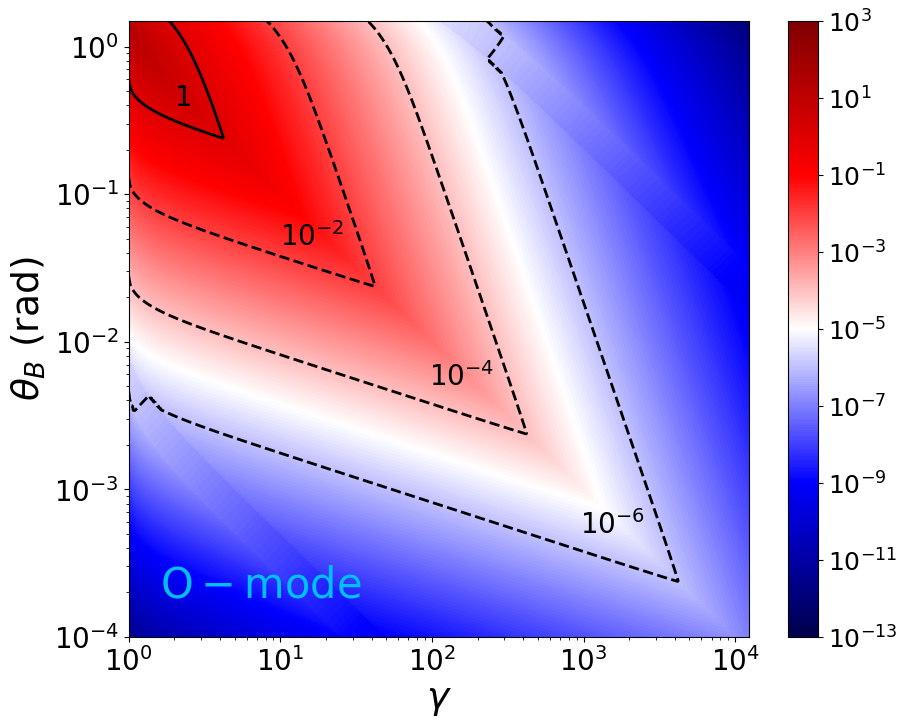}}&
\resizebox{60mm}{!}{\includegraphics[]{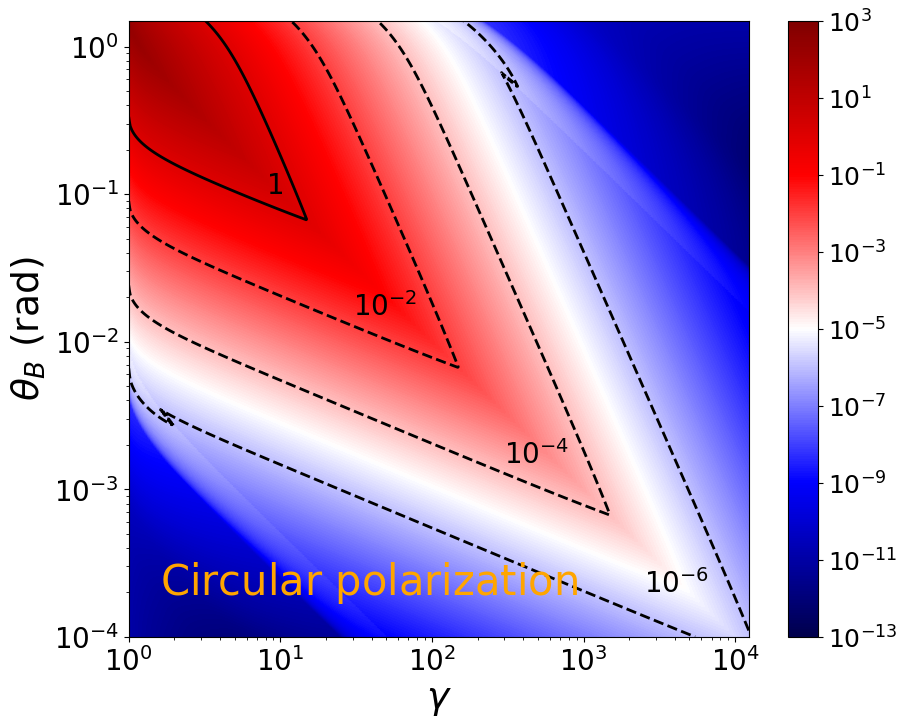}}\\
\end{tabular}
\caption{Scattering optical depth $\tau$ as a function of $\gamma$ and $\theta_B$ for X-mode (left), O-mode (middle), and circularly polarized waves (right), computed for $a_{\rm LP}=10^4$ ($a_{\rm CP}=a_{\rm LP}/\sqrt{2}$) and $\omega_B'/\omega'=35$. 
Black solid curves denote $\tau = 1$, dashed curves correspond to $\tau = 10^{-2}, 10^{-4},$ and $10^{-6}$, as labeled.
The optical depth decreases outside each contour curve, i.e., regions enclosed by a given curve correspond to optical depths larger than the value indicated on that curve.
Since $\tau \propto \xi$, these four curves are also effective $\tau = 1$ curves for three multiplicity values $\xi = 1, 10^2, 10^4, 10^6$, respectively.
The spin period of the magnetar is $P=3 \ \rm s$.}
\label{fig:tau}
\end{center}
\end{figure*}

Figure~\ref{fig:tau} shows the scattering optical depth for X-mode, O-mode, and circularly polarized waves as a function of the bulk Lorentz factor $\gamma$ and the propagation angle $\theta_B$.
The multiplicity factor $\xi$ is normalized to unity.
The contours mark $\tau=1$ (black solid), $10^{-2}, 10^{-4}, 10^{-6}$ (black dashed, as labeled), respectively.
For all polarization states, FRBs can escape the magnetosphere in the region below and to the right of these curves.
In particular, for typical FRB luminosities, transparency is achieved for $\gamma\gtrsim 10^2$, with the required Lorentz factor decreasing further as the propagation becomes increasingly aligned with the magnetic field.
These results demonstrate that relativistic particle motion and favorable geometry naturally suppress the scattering optical depth, enabling FRB escape along open magnetic field lines.

We note that a special propagation angle exists for which $\sin\theta_B=1/\gamma$ in the lab frame corresponds to $\theta_B'=\pi/2$ in the rest frame of the relativistic particle.
At this angle the scattering cross section reaches its maximum (see the left panel of Figure~\ref{fig:sigma}), leading to an enhanced optical depth.
However, this configuration represents a fine-tuning condition.
As FRB waves propagate through the magnetosphere, $\theta_B$ varies with radius, and the wave traverses this specific angle over a very short spatial and temporal interval.
It is therefore unlikely that the system remains near this configuration for a duration comparable to the FRB timescale, and we do not expect it to contribute significantly to the wave -- particle interaction.

\section{Magnetar Crust Quakes and Impacts of MHD Waves}\label{sec:open}

\subsection{Dynamical Evolution of the Magnetic Field}

Alfv\'en waves can be excited near the magnetar surface by crustal quakes \citep{Thompson&Duncan1996,Yuan2020,Yuan2022}. 
In a dipolar magnetic field, the magnetic field strength of Alfv\'en waves scales as
\begin{equation}
B_{\rm aw}=B_{\rm aw,0}\left(\frac{r}{R_\star}\right)^{-3/2},
\end{equation}
where $B_{\rm aw,0}$ is the magnetic field strength of Alfv\'en waves at the magnetar surface.
The relative wave amplitude is smaller than unity near the stellar surface and increases with radius according to
\begin{equation}
\frac{B_{\rm aw}}{B}=f_B\left(\frac{r}{R_\star}\right)^{3/2},
\end{equation}
where $f_B=B_{\rm aw,0}/B_\star$ denotes the ratio of the Alfv\'en wave amplitude to the background magnetic field at the magnetar surface.
The non-linear radius at which the Alfv\'en wave amplitude becomes comparable to the background field strength, $B_{\rm aw}=B$, is therefore given by
\begin{equation}\label{eq:R_E_aw}
R_{\rm E,aw}=f_B^{-2/3}R_\star\simeq (4.6\times10^8 \ {\rm cm}) \ f_{B,-4}R_{\star,6}<R_{\rm E,frb}.
\end{equation}
This implies that Alfv\'en waves enter the nonlinear regime before FRB waves reach $R_{\rm E,frb}$. 
As a result, the background magnetic field is already opened into a perturbed split monopole configuration which is captured by force-free electrodynamic simulations \citep{Burnaz2025}, and FRB waves propagate in a nearly quasi-parallel geometry. 
This field opening naturally suppresses non-linear scattering cross section and facilitates FRB escape.
The energy of Alfv\'en waves generated by crustal quakes is expected to peak near the magnetic pole and decrease toward the equatorial region \citep{Qu&Bransgrove}. 
As a result, FRBs triggered by crustal quakes are more likely to propagate through a larger open field line region, as illustrated in the lower panel of Figure~\ref{fig:cartoon}.

As Alfv\'en waves propagate outward, they can escape through the open field line region, where the local plasma density decreases and the waves convert into O-mode electromagnetic waves that propagate freely. 
By contrast, Alfv\'en waves entering the closed field line region are reflected and gradually lose their energy. 
Once the energy of Alfv\'en waves disappears, the magnetosphere is expected to relax toward a more dipolar configuration. 
The initial readjustment of the magnetic field occurs on a dynamical timescale comparable to the light crossing time of the perturbed region, i.e. $\delta t_{\rm aw}\sim R_{\rm E,aw}/c\sim (10 \ {\rm ms}) \ f_{B,-4}R_{\star,6}$.

\subsection{Wave -- Wave Conversion}

In the low-frequency MHD regime, $\omega \ll \omega_p$, the plasma supports two force-free eigenmodes: the fast mode and the Alfv\'en mode.
In a strongly magnetized magnetosphere, only the fast mode can freely propagate across magnetic field lines, whereas Alfv\'en waves are guided along the background field. In a curved dipolar magnetic geometry, nonlinear interactions allow for resonant mode coupling between these two branches, so that fast waves may convert into Alfv\'en waves and vice versa \citep{Thompson&Blaes1998,Lyubarsky2019,Golbraikh&Lyubarsky2023}.
However, such wave -- wave conversion critically relies on the presence of a finite transverse wavenumber with respect to the background magnetic field, $k_\perp \neq 0$. 
Physically, nonlinear interactions in the force-free plasma are mediated by currents induced by the transverse distortion of magnetic field lines. 
When $\vec k\parallel\vec B$, the Alfv\'en and fast modes become degenerate and carry no charge or current perturbations. 
As a result, efficient mode conversion is suppressed when the waves propagate nearly parallel to the background magnetic field.
This situation is particularly relevant if a large-amplitude Alfv\'en wave launched by the magnetar crustal quake significantly straightens the background magnetic field in the emission region \citep{Burnaz2025}. 
In such a scenario, even if the FRB frequency satisfies $\omega \ll \omega_p$, the radiation propagates quasi-parallel to $\vec{B}$, and wave -- wave conversion becomes ineffective.

Observational constraints strongly support this picture. 
Active repeating FRBs can exhibit extremely high circular polarization \citep{Jiang2024}, together with diverse PA swing in non-repeating FRBs and repeating FRBs, and abrupt PA jumps within millisecond timescales \citep{Luo2020nature,Mckinven2024,NiuJR2024}. 
If only a single orthogonal plasma eigenmode could propagate, one would instead expect highly linearly polarized radiation with a flat PA evolution. The observed polarization complexity therefore implies that the FRB emission is already generated in a regime where both polarization modes are available.

\section{Conclusions and Discussions}\label{sec:Conclusion}

In this paper, we have investigated the propagation of a large-amplitude electromagnetic wave, associated with a bright FRB, through the linear and non-linear regions of a magnetar magnetosphere. 
In particular, we calculated the scattering cross section and estimated the optical depth of linearly polarized X-mode, O-mode, mixed mode, and circularly polarized FRB waves interacting with the single test charged particle in a background magnetic field.
We provide the argument that magnetar crust quakes as the trigger mechanism for producing FRBs can dynamically modify the magnetospheric geometry and enhance the transparency of FRBs.
We also discuss the non-linear condition for monster shock formation.
Our main conclusions can be summarized as follows:

\begin{itemize}
\item In Section~\ref{sec:Particle motion}, we presented a general description of charged-particle motion and calculated the scattering cross sections for linearly and circularly polarized electromagnetic waves propagating in a non-zero static background magnetic field. We showed that the scattering cross sections of both X-mode and O-mode waves remain broadly consistent with the linear scattering expressions in the highly magnetized regime $\omega'_B/\omega' \gg 1$, even when the wave amplitude is large.
This occurs because the strong background magnetic field constrains the particle motion, effectively suppressing the wave induced transverse dynamics even for $a\gg1$.
\item In Section~\ref{sec:Scattering}, we investigated the propagation of the O-mode in the linear regime ($E_w < B$) inside the magnetosphere. 
For particles initially at rest, curvature radiation can in principle extract energy from O-mode FRB waves. However, we found that as long as $\theta_B \lesssim 0.1$ when $\xi=10^2$, the associated energy loss remains below the percent level. 
However, at sufficiently large $\xi$, strong energy dissipation and increased optical depth prevent FRB emission from escaping magnetar magnetospheres.
The condition for pair production is not satisfied. 
If the particles are instead initially moving relativistically with a bulk Lorentz factor $\Gamma$, as is likely in FRB-producing environments, the pair production threshold is further suppressed because the required photon energy is effectively increased by a factor of $\Gamma$. 
Despite the presence of a parallel electric field component in the O-mode, we conclude that its associated damping processes remain inefficient in the quasi-parallel regime.
In this case, we find that the resulting optical depth is much smaller than unity (Equation~(\ref{eq:tau_O_linear})).
\item In Section~\ref{sec:Numerical results}, we numerically solved the
relativistic equations of motion for an $e^\pm$ initially at rest interacting with large-amplitude linearly and circularly polarized FRB waves in the presence of a strong background magnetic field. 
We found that the averaged scattering cross sections depend sensitively on $\omega_B/\omega$, $\theta_B$, and the wave amplitude $a$. 
For X-mode waves, the effective cross section becomes much smaller than the Thomson cross section when $a \sin\theta'_B < \omega'_B/\omega'$, leading to an optical depth in the outer magnetosphere that is well below unity. 
O-mode waves generally exhibit larger cross sections due to the unscreened parallel electric field when $\theta_B > 0$, but the X- and O-mode cross sections become comparable in the limit $\theta_B \rightarrow 0$ (Figure~\ref{fig:sigma_omega_B}). 

At larger $\theta'_B$, the cross sections of both X- and O-modes follow an empirical power-law scaling $\sigma'\propto \theta_B'^{\,3.5}$ (Equations~(\ref{eq:sigma_X}) \& (\ref{eq:sigma_O})). 
The scattering cross section of linearly polarized mixed modes differs from those of pure X- and O-modes by roughly an order of magnitude. 
\item We presented in Section~\ref{sec:Numerical results} the scattering optical depth for a bright FRB with luminosity $10^{42} \ {\rm erg \ s^{-1}}$ in a magnetar magnetosphere, explicitly accounting for X-mode, O-mode, and circular polarization. 
We found that the magnetar magnetosphere is transparent to large amplitude FRB radio waves for a large region of the $\gamma_p$--$\theta_B$ parameter space for moderate multiplicities.
However, for sufficiently large multiplicities, FRBs may not escape from the magnetosphere.
\item Magnetar crust quakes are expected to launch large-amplitude Alfv\'en waves near the stellar surface. 
As these waves propagate outward in a dipole magnetic field, their relative amplitude grows and they enter the nonlinear regime at radii smaller than the characteristic emission radius of FRB waves (Equation~(\ref{eq:R_E_aw})). 
This leads to a substantial opening and straightening of the background magnetic field into a perturbed, split-monopole-like configuration \citep{Burnaz2025}. 
In addition, if the background field is transiently driven toward a monopolar-like configuration with $B\sim r^{-2}$, the ratio $\omega_B'/\omega'$ remains larger, making it more difficult for FRB waves to enter the nonlinear scattering regime. 
As a result, FRB waves propagate through the magnetosphere in a quasi-parallel geometry with respect to the local magnetic field. This effect further enhances magnetospheric transparency.
In this configuration, both nonlinear scattering and wave -- wave conversion between fast and Alfv\'en modes are strongly suppressed.
This naturally enhances the transparency of the magnetosphere to FRB emission. 
Furthermore, the observed high circular polarization, diverse PA swings and abrupt PA jumps in both repeating and non-repeating FRBs indicate that the radiation is generated in a regime where both orthogonal modes should escape.
These observational properties therefore support an emission scenario in which FRBs are produced at frequencies comparable to or exceeding the background pair plasma frequency in the comoving frame.

\end{itemize}

Recent MHD solutions of fast mode propagation show that they can steepen and form monster shocks \citep{ChenYR2022,Beloborodov2023,Vanthieghem&Levinson2025,Bernardi2025}.
This process can place constraints on FRB propagation.
We consider a linearly polarized fast mode associated with FRBs, with electric field amplitude $E$, propagating obliquely to the background magnetic field.
The fast mode enters the non-linear region when $B^2-E^2=B_{}^2-2BE\sin\theta_{B}<0$. One can see that $E\sin\theta_B$ is Lorentz invariant.
The nonlinear radius is obtained from the condition $E_w > B/(2\sin\theta_B)$, which gives $R_{\rm nl}\gtrsim(10^9 \ {\rm cm}) \ L_{\rm frb,42}^{-1/4}B_{\star,15}^{1/2}R_{\star,6}^{3/2}$, where we adopt $\theta_B=5^\circ$ for the fast mode propagation angle.
For these parameters, the quantity $\eta=\omega_p^2R_{\rm nl}/c\omega_B\simeq0.4<10$, which implies that a shock is unlikely to form \citep{ChenYR2022}.
Notably, this radius is close to the light cylinder radius, where GHz FRB waves are no longer well described by the MHD regime due to the plasma frequency decreasing significantly.
These results indicate that FRB waves can escape from the magnetosphere along open field lines without forming a shock.

FRB waves are not expected to undergo significant resonant cyclotron absorption since the resonance condition is rapidly disrupted. 
In the presence of large-amplitude waves, particles interacting with the leading edge of the wave are quickly accelerated to relativistic energies, which shift them out of cyclotron resonance and suppress sustained absorption.

We also noticed that FRBs have extremely large photon occupation numbers, thus induced Compton scattering could influence the interaction between the radiation field and the magnetospheric pair plasma \citep{Lyubarsky2008,Lu&Kumar2018}. 
This point has been stressed in recent studies of induced scattering and related parametric instabilities in magnetized pair plasma
\citep[e.g.,][]{Nishiura&Ioka2024,Nishiura2025,Nishiura&Ioka2025,Nishiura2026}. 
The strong background magnetic field can suppress the scattering of X-mode radiation.
We note that the analytic estimates in these studies are formulated for the weak electromagnetic wave regime with $a\ll 1$.
Recent work applies analytical theory and kinetic simulations to study propagation in the strong wave regime with $a>1$ in the non-magnetized plasma environment \citep{Sridhar2026} and the waves can propagate freely without significant attenuation from induced Compton scattering when $a\omega_p\ll\omega$.
A complete treatment of induced scattering in the presence of a strong wave with $a\gg1$ and a strong background magnetic field ($\omega_B\gg\omega$), remains an important problem for future work.
Motivated by these considerations, we
provide an order of magnitude estimate of induced Compton scattering for an X-mode FRB pulse propagating through a strong background magnetic field in Appendix~\ref{Appendix}.
We conclude that the opening of magnetic field lines and relativistic plasma streaming along the field are expected to significantly reduce the induced scattering cross section.

\section*{Acknowledgements}
YQ's work is supported by the Nevada Center for Astrophysics and a Top Tier Doctoral Graduate Research Assistantship (TTDGRA) at University of Nevada, Las Vegas. 
PK’s work was funded in part by a NASA grant 80NSSC24K0770.

\section*{Data Availability}
The code developed to perform the calculation in this paper is available upon request.

\bibliographystyle{mnras}
\bibliography{example}

\appendix

\section{Induced Compton Scattering of FRBs inside the magnetar magnetosphere}\label{Appendix}

In this Appendix, we estimate whether induced Compton scattering can substantially attenuate FRB waves propagating through the magnetar magnetosphere.  
Induced Compton scattering of FRBs has been discussed extensively in the literature \citep{Wilson&Rees1978,Lyubarsky2008,Lu&Kumar2018,Nishiura&Ioka2024,Nishiura2025,Nishiura&Ioka2025,Nishiura2026,Sridhar2026}. 
The estimate below should be interpreted as an order of magnitude upper limit in a static magnetosphere.

We consider an X-mode electromagnetic wave propagating through a background magnetic field $\vec B$.
In the strong wave
regime, the particle Lorentz factor is approximately scaled as $\gamma \approx a$.
The induced Compton scattering enhancement is controlled by the photon occupation number of the final photon state. 
The photon occupation number is related to the specific intensity
\begin{equation}
n_\nu=\frac{c^2 I_\nu}{2h\nu^3},
\label{eq:app_nnu}
\end{equation}
and the brightness temperature can be defined through
\begin{equation}
k_{\rm B}T_b=\frac{c^2 I_\nu}{2\nu^2}=n_\nu h\nu.
\label{eq:app_Tb_def}
\end{equation}
The dimensionless stimulated enhancement factor can therefore be written as $k_{\rm B}T_b/m_ec^2$ which represents the Bose enhancement associated with the large radio photon occupation number.

In the absence of a strong magnetic field, the Thomson differential cross section for an electron at rest is given by \citep{Rybicki&Lightman1979}
\begin{equation}
\frac{d\sigma_{\rm T}}{d\Omega}=\frac{3\sigma_{\rm T}}{16\pi}\left(1+\cos^2\theta_s\right),
\label{eq:app_thomson_diff}
\end{equation}
where \(\theta_s\) is the scattering angle. 
Only photons scattered into final directions and frequencies that remain highly occupied by the FRB radiation field receive the induced scattering enhancement. 
For a relativistic particle with the Lorentz factor $\gamma$, the angular size of the radiation beam at radius $r$ is $\theta_{b}\sim {R_e}/{\gamma r}\ll1$, where $R_e$ is the characteristic emission radius of FRBs.
Integrating Equation~(\ref{eq:app_thomson_diff}) over this small cone $\theta_b$ gives
\begin{equation}
\sigma_{\rm cone}\simeq\int_0^{\theta_{b}}\frac{3\sigma_{\rm T}}{8\pi}2\pi\theta d\theta=\frac{3\sigma_{\rm T}}{8}\theta_{b}^2.
\label{eq:app_sigma_cone_beam}
\end{equation}
One can see that only a small fraction of the scattered photons remain in the highly occupied part of the radio beam.

For an X-mode wave in a strongly magnetized plasma with $\omega\ll \omega_B$, the transverse motion of the electron is strongly suppressed by the background magnetic field. 
To leading order, the scattering cross section scales as $\sigma_{\rm X}\simeq\sigma_{\rm T}
({\omega}/{\omega_B})^2$.
For an order of magnitude estimate, we assume that the dominant effect of the strong magnetic field is to multiply the ordinary Thomson angular pattern by the X-mode suppression factor in Equation~(\ref{eq:app_sigma_cone_beam}) and we have the effective cross section 
\begin{equation}
\sigma_{\rm eff}\simeq\frac{3\sigma_{\rm T}}{8}\theta_b^2\left(\frac{\omega}{\omega_B}\right)^2.
\label{eq:app_sigma_eff_gamma}
\end{equation}
The induced Compton effective cross section can then be estimated as $\sigma_{\rm ind}\approx\sigma_{\rm eff}({k_{\rm B}T_b}/{m_ec^2})$,
where $\sigma_{\rm eff}$ is the single particle X-mode cross section into the relevant final photon states, and $k_{\rm B}T_b/m_ec^2$ is the stimulated enhancement factor.
It should be pointed out that the FRB spectra has a finite bandwidth (FWHM) $\Delta\nu=f_\nu\nu$, where $f_\nu=\Delta\nu/\nu$ is the fractional bandwidth.
Repeating FRBs have a narrow band with a typical value $f_\nu\sim0.1-0.3$ \citep{PKumar&Shannon2021,ZhouDJ2022,ZhangYK2023,ZhouDJ2025}. 
The spectral flux is $F_\nu\approx {F}/{\Delta\nu}={L_{\rm FRB}}/{4\pi r^2 f_\nu\nu}\sim I_\nu$.
Thus we obtain
\begin{equation}
\frac{k_{\rm B}T_b}{m_ec^2}\approx\frac{c^2 I_\nu}{2\nu^2m_ec^2}=\frac{\pi^2 L_{\rm FRB}}{m_e r^2 f_\nu \omega^3},
\label{eq:app_Tb_nu}
\end{equation}
and the cross section of the induced Compton scattering can be written as
\begin{equation}
\sigma_{\rm ind}\approx\frac{3\pi^2}{8}\frac{L_{\rm FRB}}{m_e r^2 f_\nu \omega^3}\theta_b^2\left(\frac{\omega}{\omega_B}\right)^2\sigma_{\rm T}.
\label{eq:app_sigma_ind_gamma}
\end{equation}
For radiation confined to a narrow beam with $a\theta_b\ll1$, the strong wave correction scales approximately as $a^{-1}$ because the effective mass of electrons is increased by a factor $a$ thereby reducing the electron recoil by that amount, we have
\begin{equation}
\begin{aligned}
{\sigma_{\rm ind}}&\approx
\frac{3\pi^2}{8}
\frac{L_{\rm FRB}}{m_er^2f_\nu\omega^3}
\theta_b^2
\left(\frac{\omega}{\omega_B}\right)^2
\frac{1}{a}\sigma_{\rm T}\\
&=\frac{3\pi^2}{8}
\frac{1}{f_\nu}
\frac{m_e^2c^{7/2}L_{\rm FRB}^{1/2}R_e^2r^3}{e^3\gamma^2B_\star^2R_\star^6}\sigma_{\rm T}
\end{aligned}
\end{equation}
for a dipole background magnetic field. Note that if radio waves are produced at radius $R_e$ by a source with Lorentz factor $\gamma$ then the transverse size of the source is $\sim R_e/\gamma$, and its angular size at radius $r$ is $\theta_b\sim R_e/(\gamma r)$.
Thus the induced Compton scattering optical depth for non-relativistic plasma can be estimated as
\begin{equation}
\begin{aligned}
\tau_{\rm ind}&\simeq \xi n_{\rm GJ}\sigma_{\rm ind}r\\
&\simeq6.0\times10^{-3} \ \xi_5 L_{\rm frb,42}^{1/2}r_8R_{e,7}^2\gamma_2^{-2}B_{\star,15}^{-2}R_{\star,6}^{-6}\ll1
\end{aligned}
\end{equation}
for a pair multiplicity $\xi=10^5$.
One can see that the magnetosphere is optically thin to induced Compton scattering even without invoking relativistic plasma outflow or field line region opening.
We consider that the pair plasma is relativistically moving in the open field line region, the cross section is transformed to the lab frame as
\begin{equation}
\begin{aligned}
\sigma_{\rm ind}&=\frac{3\pi^2}{8}
\frac{1}{f_\nu}
\frac{m_e^2c^{7/2}L_{\rm FRB}^{1/2}R_e^2r^3}{e^3\gamma^2B_\star^2R_\star^6}(1-\beta\cos\theta_B)\sigma_{\rm T}\\
&\simeq\frac{3\pi^2}{16}
\frac{1}{f_\nu}
\frac{m_e^2c^{7/2}L_{\rm FRB}^{1/2}R_e^2r^3}{e^3\gamma^4B_\star^2R_\star^6}\sigma_{\rm T},
\label{eq:app_sigma_ind_dipole}
\end{aligned}
\end{equation}
where $1-\beta\cos\theta_B\approx1-\beta\approx1/2\gamma^2$ is applied for $\gamma\gg1$ and $\theta_B\approx 0$.
One can see that $\sigma_{\rm ind}\propto \gamma^{-4} r^3$, the static dipole estimate is strongly suppressed when the background plasma is relativistically moving with $\gamma\gg1$ at small radii.

We also evaluate the importance of induced Compton scattering of waves in a  direction outside the FRB beam. 
The occupation number of photon states corresponding to wave-vectors different from the FRB photons is small initially. However, it grows exponentially due to induced Compton scattering instability. 
For strongly magnetized plasma, 
the growth rate of this instability for photons scattered outside the beam is given by
\begin{equation}
\begin{aligned}
\Gamma_{\rm grow}&\simeq\frac{\omega \omega_p^2 a^2}{8\omega_B^2}\simeq(9.1\times10^{-2} \ {\rm s^{-1}}) \ \xi L_{\rm FRB,42}^{1/2}r_8^2P^{-1}B_{\star,15}^{-1}R_{\star,6}^{-3},
\end{aligned}
\end{equation}
where $\omega_p=(4\pi e^2 n/\gamma m_e)^{1/2}$ and we take $\gamma\sim a$.
The corresponding number of e-foldings over the escape time $t_{\rm esc}\sim r/c$ is
\begin{equation}
\Gamma_{\rm grow}t_{\rm esc}
\simeq
3.0\times10^{-4} \ \xi L_{\rm FRB,42}^{1/2}r_8^3P^{-1}B_{\star,15}^{-1}R_{\star,6}^{-3}\ll1.
\label{eq:app_growth_efold}
\end{equation}
For a low multiplicity plasma in the open field line region, off-beam induced Compton scatterings do not seem to substantially deplete the FRB pulse energy.

We conclude that several effects could significantly modify the induced Compton scattering cross section inside the magnetosphere in the following: 
(i) The magnetic field lines are expected to open up due to the propagation of Alfv\'en waves excited by magnetar crust quakes (see Section~\ref{sec:open} for a discussion). 
In this case, the field geometry may be closer to an open, approximately monopolar configuration rather than a dipole field. 
Because a monopole like field decreases more slowly with radius than a dipole field, the local background magnetic field in the outer magnetosphere can be larger than the dipole estimate.
(ii) The plasma can stream relativistically with a bulk Lorentz factor $\gamma\gg1$ along the magnetic field, which can reduce the cross section.
(iii) In the strong wave regime, the FRB wave can efficiently accelerate particles to relativistic speeds. 
The particle motion is then determined by the incoming FRB wave and the background magnetic field, and the particle velocity cannot remain aligned with the FRB propagation direction.
This can modify the scattering kinematics and shift the scattered photon frequency outside the narrow observed FRB bandwidth, thereby reducing the final state photon occupation number and weakening the induced enhancement.
These effects can substantially reduce the optical depth of induced Compton scattering inside the magnetosphere.

\bsp	
\label{lastpage}
\end{document}